\documentclass[12pt,peerreview,onecolumn]{IEEEtran}

\IEEEoverridecommandlockouts

\usepackage{color}
\usepackage{graphicx}
\usepackage{amsmath}
\usepackage{amssymb}
\usepackage{algorithm}
\usepackage{algorithmic}
\usepackage{amsmath}
\usepackage{multirow}
\usepackage{booktabs}
\usepackage{array}
\usepackage{amsthm}
\usepackage{lipsum}
\usepackage{enumerate}
\usepackage{stfloats}

\newcommand{\be}{\begin{equation}}
\newcommand{\ee}{\end{equation}}
\newcommand{\bea}{\begin{eqnarray}}
\newcommand{\eea}{\end{eqnarray}}
\newcommand{\ba}{\begin{array}}
\newcommand{\ea}{\end{array}}
\newcommand{\nid}{\noindent}
\newcommand{\non}{\nonumber}

\newcommand{\tabincell}[2]{\begin{tabular}{@{}#1@{}}#2\end{tabular}}

\title{Hybrid Beamforming with Dynamic Subarrays and Low-resolution PSs for mmWave MU-MISO Systems \thanks{This paper is supported by the National Natural Science Foundation of China (Grant No. 61671101, 61601080, and 61761136019).}}

\author{Hongyu Li,~\IEEEmembership{Student Member,~IEEE,}
        Ming Li,~\IEEEmembership{Senior Member,~IEEE,}
        and Qian Liu,~\IEEEmembership{Member,~IEEE,}

\thanks{H. Li and M. Li are with the School of Information and Communication Engineering, Dalian University of Technology, Dalian 116024, China, (e-mail: hongyuli@mail.dlut.edu.cn, mli@dlut.edu.cn).}
\thanks{Q. Liu is with the School of Computer Science and Technology, Dalian University of Technology, Dalian 116024, China, (e-mail: qianliu@dlut.edu.cn).}
}

\begin{document}
\maketitle
\vspace{-0.8 cm}
\begin{abstract}
Analog/digital hybrid beamforming architectures with large-scale antenna arrays have been widely considered in millimeter wave (mmWave) communication systems because they can address the tradeoff between performance and hardware efficiency compared with traditional fully-digital beamforming.
Most of the prior work on hybrid beamforming focused on fully-connected architecture or partially-connected scheme with fixed-subarrays, in which the analog beamformers are usually realized by infinite-resolution phase shifters (PSs).
In this paper, we introduce a novel hybrid beamforming architecture with \textit{dynamic} subarrays and hardware-efficient \textit{low-resolution} PSs for mmWave multiuser multiple-input single-output (MU-MISO) systems.
By dynamically connecting each RF chain to a non-overlap subarray via a switch network and PSs, we can exploit multiple-antenna and multiuser diversities to mitigate the performance loss due to the use of practical low-resolution PSs.
An iterative hybrid beamformer design algorithm is first proposed based on fractional programming (FP), aiming at maximizing the sum-rate performance of the MU-MISO system. In an effort to reduce the complexity, we also present a simple heuristic hybrid beamformer design algorithm for the dynamic subarray scheme.
Extensive simulation results demonstrate the advantages of the proposed hybrid beamforming architecture with dynamic subarrays and low-resolution PSs compared to existing fixed-subarray schemes.
\end{abstract}

\vspace{-0.7 cm}


\maketitle

\section{Introduction}

Millimeter wave (mmWave) communications have been deemed as one of key enabling technologies in 5G and beyond networks because of the significant advantages of providing multi-gigahertz frequency bandwidth and high data rate \cite{Pi CM 11}, \cite{Rappaport IA 13}.
Thanks to small antenna size at mmWave frequencies, a large-scale antenna array can be packed into a small area.
This feature facilitates the implementation of massive multiple-input multiple-output (MIMO) in mmWave systems, which can provide sufficient beamforming gains to combat the severe path loss in mmWave channels \cite{Swindlehurst 14}, \cite{Heath 16}.
Nevertheless, realizing the beamforming with large-scale antenna arrays is not straightforward in mmWave systems.
Conventional fully-digital beamforming architecture needs to equip one dedicated radio-frequency (RF) chain  (including analog-to-digital/digital-to-analog converter (ADC/DAC), etc.) to each antenna.
Unfortunately, this fully-digital beamforming architecture cannot be practically employed in mmWave massive MIMO systems due to the unaffordable cost and power consumption of large numbers of mmWave RF chains and other hardware components \cite{Heath 2014}, \cite{Poon 12}.
Recently, analog/digital hybrid beamforming has been advocated as a practical solution for mmWave massive MIMO systems to balance the system performance and hardware efficiency \cite{Han 2015}. The hybrid beamforming architecture employs only a few expensive RF chains to realize low-dimensional digital beamformers and utilizes a large number of cost-efficient phase shifters (PSs) to implement high-dimensional analog beamformers.
While the analog beamformers in the RF domain can provide sufficient beamforming gain to compensate for the huge path loss of mmWave channel, the digital beamformers in the baseband domain are able to offer the flexibility to realize multiuser/multiplexing techniques.
Because of its efficiency and effectiveness in mmWave systems, hybrid beamforming has attracted extensive attention from both academia and industry in recent years.

\subsection{Prior Work}

\begin{figure}[!t]
\centering
\includegraphics[width= 3.3 in]{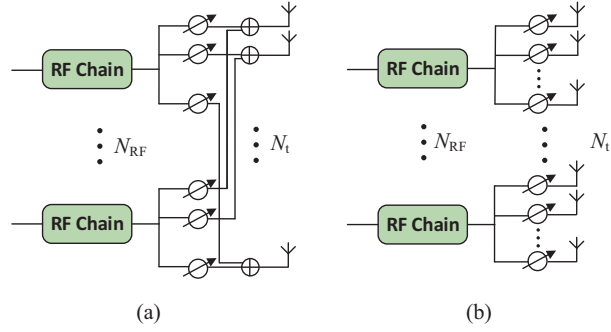}
\vspace{-0.5 cm}
\caption{Traditional hybrid beamforming architectures: (a) fully-connected; (b) partially-connected (fixed-subarray without overlapping). }\label{fig:traditional_connections}\vspace{-0.5 cm}
\end{figure}

Two typical hybrid beamforming architectures have been widely considered for mmWave MIMO systems, i.e. fully-connected architecture as shown in Fig. \ref{fig:traditional_connections}(a) and partially-connected
architecture as shown in Fig. \ref{fig:traditional_connections}(b).
In the fully-connected hybrid beamforming structure, each RF chain is connected to all antennas via an analog network of PSs.
The challenge in designing such hybrid beamformer mainly lays in the practical constraints of PSs, such as constant modulus.
In point-to-point mmWave MIMO systems, maximizing spectral efficiency is often approximated by minimizing the Euclidean distance between the hybrid beamformer and the fully-digital beamformer \cite{Yu 2016}-\cite{Ni TSP 17}.
Codebook-based hybrid beamformer designs are also widely used \cite{Ayach TWC 14}-\cite{He TSP 17}, in which the analog beamformers are picked from certain candidate vectors, such as array response vectors and
discrete Fourier transform (DFT) beamformers.
Hybrid beamformer design for mmWave multiuser systems have also been investigate in \cite{Li 2017}-\cite{ZH 2017}.
While fully-connected hybrid beamforming enjoys satisfactory spectral efficiency performance close to that obtained with fully-digital solutions, it still suffers from relatively high power consumption and hardware complexity due to the use of a large number of PSs.
Therefore, partially-connected architecture \cite{Heath 13}-\cite{Huang WCL}, in which each RF chain is connected to a fixed disjoint subset of antennas, has been proposed to reduce the number of PSs and further decrease the power consumption and hardware complexity.

The aforementioned hybrid beamformer designs generally assume the use of \textit{infinite/high-resolution} PSs, which require complicated hardware
circuits and have high energy consumption at mmWave frequencies \cite{Poon 12}.
Thus, other than reducing the number of PSs, using cost-effective and energy-saving \textit{low-resolution} PSs to implement analog beamformers is another hardware-efficient approach \cite{Li WCM}.
Several literatures have investigated the hybrid beamformer design with finite/low-resolution PSs for the fully-connected architecture \cite{Yu LR-PS}-\cite{Zhang TVT 2018}.
In order to achieve the maximum hardware efficiency, \cite{Gao ICC 17}, \cite{ZH C2018} investigated partially-connected architectures with 1-bit (i.e. binary) PSs for a point-to-point mmWave MIMO system.
However, the performance of traditional partially-connected schemes with fixed-subarray and low-resolution PSs is not always satisfactory due to the following two reasons: \textit{i}) The employment of low-resolution PSs makes it difficult to finely control the beam and thus will cause notable performance degradation; \textit{ii)} fixed-subarray architecture limits the flexibility of large antenna arrays and further influences beamforming accuracy.
Therefore, a meaningful research direction is seeking efficient solutions to tackle above two problems for partially-connected hybrid beamforming architectures with low-resolution PSs.

Recently, dynamic connection/mapping strategy, which adaptively partitions all transmit antennas into several subarrays associated with different RF chains, has been proposed to mitigate the performance degradation caused by fixed-subarrays.
Several literatures have demonstrated the benefits on both spectral efficiency and energy efficiency by applying the dynamic connection strategy in the hybrid beamforming with high/finite resolution PSs \cite{Heath 17}-\cite{Hongyu VTC 2019}.
Moreover, recent works \cite{Hongyu CL 2018}, \cite{Hongyu TVT 2019} illustrate that adaptively selecting a subset of transmit antennas form a large-scale antenna array can exploit the augmented antenna diversity to effectively compensate for the accuracy loss of low-resolution PSs.
Motivated by these findings, we attempt to take full advantage of both the flexibility of the dynamic antenna scheme and the multiple-antenna diversity of large-scale antenna arrays to enhance the performance of the partially-connected hybrid beamforming with low-resolution PSs.


\subsection{Contributions}

In this paper, we consider the problem of the hybrid beamformer design for mmWave multiuser multiple-input single-output (MU-MISO) systems. The contributions of this paper are summarized as follows:
\begin{itemize}
\item We introduce an efficient hybrid beamforming architecture with dynamic subarray and low-resolution PSs. In an effort to mitigate the performance loss due to the use of low-resolution PSs, each RF chain is dynamically connected to a disjoint subset of the total transmit antennas to exploit the multiple-antenna and multiuser diversities.
\item With the aid of fractional programming (FP), an effective hybrid beamformer design algorithm is proposed to maximize the sum-rate performance of the mmWave MU-MISO system.
\item In order to reduce the time complexity, we also develop a simple heuristic hybrid beamformer design algorithm to alternatively update the analog beamformer and digital beamformer until a convergence solution is obtained.
\item The time complexities of the proposed FP-based and heuristic hybrid beamformer designs are analyzed and compared.
\item The effectiveness of two proposed hybrid beamformer designs is validated by extensive simulation results, which illustrate that both two algorithms can remarkably outperform traditional fixed-subarray schemes.
\end{itemize}

\subsection{Notations:}
The following notations are used throughout this paper.
$\mathbf{a}$ and $\mathbf{A}$ indicate column vectors and matrices, respectively.
$(\cdot)^*$ denotes the conjugate operation of a complex number.
$(\cdot)^T$£¬$(\cdot)^H$ and $(\cdot)^{-1}$ denote the transpose, conjugate-transpose operation, and inversion for a matrix, respectively.
$\mathbb{E} \{ \cdot \}$ represents statistical expectation.
$\Re \{ \cdot \}$ extracts the real part of a complex number.
$\mathbf{I}_L$ indicates an $L \times L$ identity matrix.
$\mathbb{C}$ denotes the set of complex numbers.
$| \mathcal{A} |$  denotes the cardinality of set $\mathcal{A}$.
$\| \mathbf{A} \|_F$ denotes the Frobenius norm of matrix $\mathbf{A}$.
$\| \mathbf{a} \|_0$ is the 0-norm of vector $\mathbf{a}$.
Finally, $\mathbf{A}(i,:)$ and $\mathbf{A}(i,j)$ denote the $i$-th row and $(i,j)$-th element of matrix $\mathbf{A}$, respectively.

\section{System Model and Problem Formulation}

\subsection{System Model}

\begin{figure*}
\centering
\includegraphics[height=2.5in]{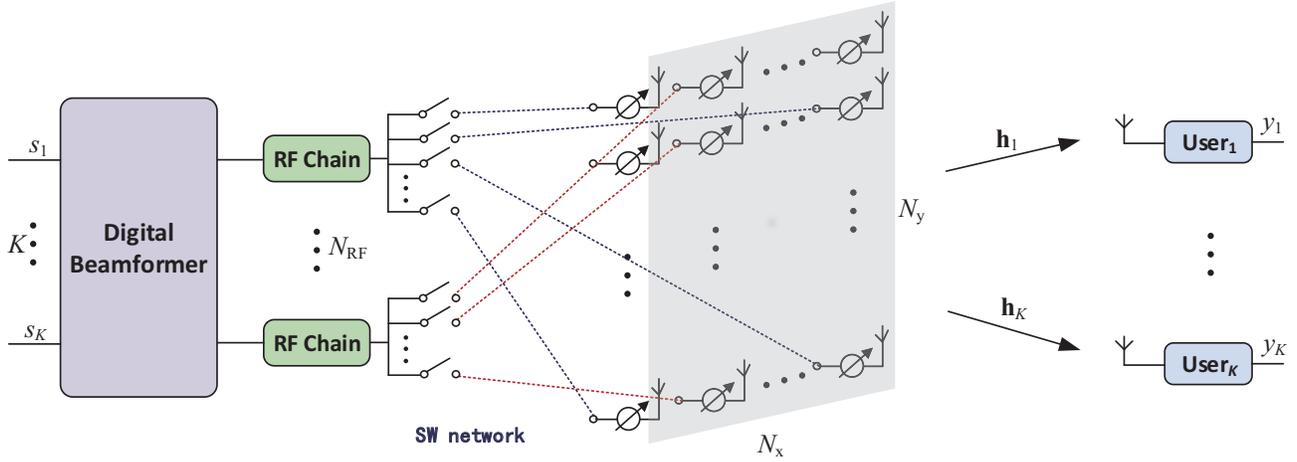}
\caption{Hybrid beamforming with dynamic subarrays and low-resolution PSs for mmWave MU-MISO system.}\label{fig:HBF} \vspace{-0.2 cm}
\end{figure*}

We consider a mmWave downlink MU-MISO system as illustrated in Fig. \ref{fig:HBF}. The base station (BS) employs a uniform plane array (UPA) with $N_\textrm{x}$ antennas in horizontal direction and $N_\textrm{y}$ antennas in vertical direction.
The total number of antennas is $N_\textrm{t} = N_\textrm{x} \times N_\textrm{y}$ and the number of RF chains is $N_\textrm{RF}$, $N_\textrm{RF} \ll N_\textrm{t}$.
By the hybrid beamforming technology, the BS can simultaneously communicate with $K$ single-antenna users.
In this paper, we assume the number of RF chains at the BS is greater than or equal to the number of users, i.e. $N_\textrm{RF} \ge K$.
The information symbols of $K$ users are firstly precoded by a digital beamforming matrix $\mathbf{F}_\textrm{BB} \triangleq [\mathbf{f}_{\textrm{BB},1}, \ldots, \mathbf{f}_{\textrm{BB},K}]\in \mathbb{C}^{N_\textrm{RF}\times K}$.
After being up-converted to the RF domain via $N_\textrm{RF}$ RF chains, the signals are further precoded in the RF domain by $N_\textrm{t}$ low-resolution PSs.
Particularly, each RF chain will be dynamically connected to a disjoint set of antennas via a switch (SW) network and corresponding PSs.
Several previous works \cite{WCL Heath 2016}-\cite{Rial Access 16} have demonstrated the feasibility and effectiveness of employing switches in hybrid beamforming architectures for mmWave MIMO systems.
By mapping each digitally-precoded data stream to one dynamic antenna subarray, the analog precoding is carried out using a set of corresponding low-resolution PSs, each of which has a constant magnitude $\frac{1}{\sqrt{N_\mathrm{t}}}$ and quantized phases controlled by the number of bits $B$.
The possible phases of each PS are within a set $\mathcal{F} \triangleq \{\frac{1}{\sqrt{N_\mathrm{t}}}e^{j\frac{2\pi b}{2^B}} | b = 0,1,\ldots,2^B-1\}$.
Obviously, the beamforming gains for each user can be easily adjusted by changing the sizes of subarrays.
More importantly, by dynamically selecting antennas and tuning the phases of the associated PSs, the multi-antenna and multiuser diversities can be utilized to appropriately compensate the accuracy loss due to the use of low-resolution PSs.
Therefore, the signal of each RF chain can dynamically select the optimal antenna subarray and employ the optimal corresponding analog beamformer according to the instantaneous channel state information (CSI) of all users, in such a way to enhance the downlink multiuser transmission performance.

To facilitate the dynamic subarray analog beamformer design, we define the analog beamformer matrix as $\mathbf{F}_{\textrm{RF}} \in \{\mathcal{F},0\}^{N_\textrm{t} \times N_\textrm{RF}}$.
Specifically, if the $k$-th RF chain is connected to the $i$-th antenna via a low-resolution PS, then the corresponding element of analog beamformer $\mathbf{F}_{\textrm{RF}}(i,k) \in \mathcal{F}$ has a nonzero phase value; otherwise, $\mathbf{F}_{\textrm{RF}}(i,k) = 0$.
To guarantee no overlap among different subarrays, the analog beamformer has only one nonzero element in each row, i.e.
$\|\mathbf{F}_{\textrm{RF}}(i,:)\|_0 = 1, i = 1,\ldots, N_\textrm{t}$.

Then, the transmitted signal at the BS can be expressed as
\begin{equation}
\mathbf{x} = \sum_{k = 1}^{K} \mathbf{F}_{\textrm{RF}} \mathbf{f}_{\textrm{BB},k} s_k,
\end{equation}
where $s_k \in \mathbb{C}, k = 1, \ldots, K$, is the transmit symbol for the $k$-th user. All the information symbols are independent within a user stream and across the users, $\mathbb{E}\{|s_k|^2\} = 1$.

Consider a narrow-band system, the received signal of the $k$-th user can be modeled as
\begin{equation}
y_k = \mathbf{h}_k^H \sum_{k = 1}^{K} \mathbf{F}_{\textrm{RF}} \mathbf{f}_{\textrm{BB},k} s_k + n_k, \label{eq:received signal 1}
\end{equation}
where $\mathbf{h}_k \in \mathbb{C}^{N_\textrm{t}}, k = 1, \ldots, K$, denotes the channel vector between the BS and the $k$-th user and
$n_k \sim \mathcal{CN} \left(0,\sigma_k^2 \right), k = 1, \ldots, K$, is the independent and identically distributed (i.i.d.) complex Gaussian noise with variance $\sigma_k^2$.

\subsection{Channel Model}

As many literatures claimed \cite{Swindlehurst 14}-\cite{Heath 2014}, \cite{Ayach TWC 14}, \cite{Rappaport 15}, mmWave channel is expected to have limited scattering due to the highly directional propagation characteristic at the mmWave frequencies. Therefore, the mmWave channel can be simply modeled as a sum of multiple propagation paths. Under this classic model, the $k$-th channel vector can be formulated as
\begin{equation}
\mathbf{h}_k = \sqrt{\frac{N_{\rm t}}{L_k}} \sum\limits_{l=1}^{L_k} \alpha_{k,l} \mathbf{a}(\phi_{k,l}, \theta_{k,l}),\label{eq:channel model}
\end{equation}
\nid where $\sqrt{\frac{N_{\rm t}}{L_k}}$ is a normalization factor, $L_k$ is the number of paths and $\alpha_{k,l}\sim \mathcal{CN}(0,1)$ is the complex gain of the $l$-th path for the $k$-th user.
$\mathbf{a}(\phi_{k,l}, \theta_{k,l})$ denotes the transmit steering vector for the $k$-th user of the $l$-th path with horizontal and vertical angles of departure (AoD) of $\phi_{k,l}$ and $\theta_{k,l}$, respectively.
For an $N_\textrm{x} \times N_\textrm{y}$ UPA with $N_{\rm t}$ total transmit antennas and inter-antenna space $d$, the transmit steering vector of the $l$-th path for the $k$-th user is given by \cite{Gao JSAC 16}
\begin{equation}
\mathbf{a}(\phi_{k,l},\theta_{k,l}) = \mathbf{a}_{\textrm{x}}(\phi_{k,l},\theta_{k,l}) \otimes \mathbf{a}_{\textrm{y}}(\theta_{k,l}),
\label{eq:AOD}
\end{equation}
where $\otimes$ is the Kronecker-product operation, $\mathbf{a}_{\textrm{y}}(\theta_{k,l})$ denotes the vertical steering vector of the $l$-th path for the $k$-th user, which has a form of
\begin{equation}
\mathbf{a}_{\textrm{y}}(\theta_{k,l}) = \frac{1}{\sqrt{N_{\rm y}}} [1, {e}^{j\frac{2\pi}{\lambda}d\cos\theta_{k,l}}, \ldots , {e}^{j(N_{\rm y}-1)\frac{2\pi}{\lambda}d\cos\theta_{k,l}}]^{T},
\label{eq:AOD1}
\end{equation}
where $\lambda$ denotes the wavelength of the signal. $\mathbf{a}_{\textrm{x}}(\phi_{k,l},\theta_{k,l})$ denotes the horizontal steering vector of the $l$-th path for the $k$-th user, which is given by
\begin{equation}
\mathbf{a}_{\textrm{x}}(\phi_{k,l},\theta_{k,l}) =  \frac{1}{\sqrt{N_{\rm x}}} [1, {e}^{j\frac{2\pi}{\lambda}d\sin\theta_{k,l}\sin\phi_{k,l}}, \ldots,
{e}^{j(N_{\rm x}-1)\frac{2\pi}{\lambda}d\sin\theta_{k,l}\sin\phi_{k,l}}]^{T}.
\label{eq:AOD2}
\end{equation}
In this paper, we assume the knowledge of $\mathbf{h}_k$ for each user is known perfectly and instantaneously to the BS, which can be obtained by the channel estimation with uplink training \cite{Heath 16}, \cite{Rial Access 16}.

\subsection{Problem Formulation}

The sum-rate of the MU-MISO downlink system is given by
\begin{equation}
R = \sum_{k = 1}^{K} \log_2(1 + \textrm{SINR}_k),
\label{eq:sum-rate}
\end{equation}
where $\textrm{SINR}_k$ is the signal-to-interference-plus-noise ratio (SINR) of the $k$-th user, which can be written as
\begin{equation}
\textrm{SINR}_k = \frac{|\mathbf{h}_k^H \mathbf{F}_{\textrm{RF}}\mathbf{f}_{\textrm{BB},k}|^2}{\sum_{j \ne k} |\mathbf{h}_k^H \mathbf{F}_{\textrm{RF}}\mathbf{f}_{\textrm{BB},j}|^2 + \sigma_k^2}, ~~~\forall k.
\end{equation}

The objective of this paper is to jointly design the digital beamformer $\mathbf{F}_\textrm{BB}$ and analog beamformer $\mathbf{F}_\textrm{RF}$ to maximize the sum-rate of the MU-MISO downlink system, subject to the following constraints: \textit{i)} constant amplitude and discrete phases of PSs, i.e. $\mathbf{F}_\textrm{RF}(i,j) \in \{0,\mathcal{F}\}, i = 1, \ldots, N_\textrm{t}, j = 1, \ldots, N_\textrm{RF}$; \textit{ii)} non-overlapping dynamic mapping, i.e. $\|\mathbf{F}_\textrm{RF}(i,:)\|_0 = 1, i = 1, \ldots, N_\textrm{t} $;  \textit{iii)} transmit power constraint, i.e. $\|\mathbf{F}_\textrm{RF}\mathbf{F}_\textrm{BB}\|_F^2 = P$, where $P$ is the transmit power.
With above constraints, the problem can be formulated as
\begin{equation}
\begin{aligned}
\label{eq:optimization problem}
\{\mathbf{F}_\textrm{RF}^{\star}, \mathbf{F}_\textrm{BB}^{\star}\} &~=~\arg \max ~~ R \\
{\rm s.t.} ~~ &\mathbf{F}_\textrm{RF}(i,j) \in \{0,\mathcal{F}\}, \forall i,j, \\
&\|\mathbf{F}_\textrm{RF}(i,:)\|_0 = 1, \forall i, \\
&\|\mathbf{F}_\textrm{RF}\mathbf{F}_\textrm{BB}\|_F^2 = P.
\end{aligned}
\end{equation}

Obviously, the optimization problem (\ref{eq:optimization problem}) is non-convex and difficult to solve due to the discrete phases of low-resolution PSs and the $l_0$ norm constraint. To tackle the difficulties in (\ref{eq:optimization problem}), we first propose an iterative design algorithm with satisfactory performance based on the theory of FP, then we develope a simple heuristic method to further reduce the time complexity.

\section{FP-based Hybrid Beamformer Design}
\label{sc:FP-based Design}

We start by analyzing the characteristic of the objective in (\ref{eq:sum-rate}). The objective is computed as $\sum_{k=1}^K \log(1 + \textrm{SINR}_k)$, which is a typical function of multiple fractional parameters (i.e. SINRs).
Motivated by the findings in \cite{WY PF I}, \cite{WY PF II} which can use FP to solve this multiple-ratio problem (such as fully-digital beamforming and power allocation problem in multiuser MIMO systems), we attempt to equivalently transform the original problem to a solvable form. Then, with the newly transformed objective function, we propose to iteratively design the analog beamformer and digital beamformer.

\subsection{Transformation of objective function}

We begin with taking the ratio parts $\textrm{SINR}_k, k = 1, \ldots, K,$ out of the logarithm. Based on the \textit{Lagrangian Dual Transform}  \cite{WY PF II}, we present the following proposition.

\vspace{0.3 cm}
\nid\textbf{Proposition 1.}
The objective function in (\ref{eq:optimization problem}) is equivalent to
\begin{equation}
f_r(\mathbf{F}_{\textrm{RF}}, \mathbf{F}_{\textrm{BB}}, \mathbf{r}) =  \sum_{k = 1}^{K} \log_2(1 + r_k) - \sum_{k = 1}^K r_k
+ \sum_{k = 1}^K\underbrace{\frac{(1 + r_k)|\mathbf{h}_k^H \mathbf{F}_{\textrm{RF}} \mathbf{f}_{\textrm{BB},k}|^2}{\sum_{j = 1}^K |\mathbf{h}_k^H \mathbf{F}_{\textrm{RF}} \mathbf{f}_{\textrm{BB},j}|^2 + \sigma_k^2}}_{\textrm{fractional term}},
\label{eq:f_r}
\end{equation}
when each element of the auxiliary variable vector $\mathbf{r} \triangleq [r_1, \ldots, r_K]^T$ has the following optimal value:
\begin{equation}
r_k^{\star} = \frac{|\mathbf{h}_k^H \mathbf{F}_{\textrm{RF}} \mathbf{f}_{\textrm{BB},k}|^2}{\sum_{j \ne k} |\mathbf{h}_k^H \mathbf{F}_{\textrm{RF}}\mathbf{f}_{\textrm{BB},j}|^2 + \sigma_k^2}, \forall k.
\label{eq:r_k}
\end{equation}

\begin{IEEEproof}
See Appendix A.
\end{IEEEproof}

By Proposition 1, with the same constraints, the problems of (9) and maximizing (10) are equivalent in the sense that $\{\mathbf{F}_{\textrm{RF}}, \mathbf{F}_{\textrm{BB}}\}$ are solutions to (9) if and only if they are also solutions to maximize (10), i.e. $\{\mathbf{F}_{\textrm{RF}}^\star, \mathbf{F}_{\textrm{BB}}^\star\}$ are common optimal solutions of these two problems.
Therefore, the original objective in (\ref{eq:sum-rate}) can be firstly transformed into an equivalent function  $f_r(\mathbf{F}_{\textrm{RF}}, \mathbf{F}_{\textrm{BB}}, \mathbf{r})$ as shown in (\ref{eq:f_r}).
Unfortunately, the hybrid beamformer design problem is still intractable due to the complicated form of the sum of $K$ fractional term (the last term as inclosed in (\ref{eq:f_r})).
Next, in an effort to facilitate the analog beamformer and digital beamformer design for fixed $r_k^{\star}, k = 1, \ldots, K$, we attempt to apply \textit{Quadratic Transform} \cite{WY PF II} on the fractional term. To achieve this goal, we develop the following proposition.

\vspace{0.3 cm}
\nid\textbf{Proposition 2.}
The fractional term in objective function (\ref{eq:f_r})
\begin{equation}
\frac{(1 + r_k)|\mathbf{h}_k^H \mathbf{F}_{\textrm{RF}} \mathbf{f}_{\textrm{BB},k}|^2}{\sum_{j = 1}^K |\mathbf{h}_k^H \mathbf{F}_{\textrm{RF}} \mathbf{f}_{\textrm{BB},j}|^2 + \sigma_k^2}, \forall k, \label{eq:SINR_r}
\end{equation}
is equivalent to
\begin{equation}
2 \sqrt{1 + r_k} \Re \{ t_k^*\mathbf{h}_k^H \mathbf{F}_{\textrm{RF}} \mathbf{f}_{\textrm{BB},k}\} - |t_k|^2\mathbf{C}_k, \forall k,
\label{eq:SINR_q}
\end{equation}
when the auxiliary variables $t_k, k = 1, \ldots, K,$ have the following optimal values:
\begin{equation}
t_k^{\star} = \frac{\sqrt{1 + r_k}\mathbf{h}_k^H \mathbf{F}_{\textrm{RF}} \mathbf{f}_{\textrm{BB},k}}{\mathbf{C}_k}, \forall k,\label{eq:y_k}
\end{equation}
where
\begin{equation}
\mathbf{C}_k \triangleq \sum_{j = 1}^K |\mathbf{h}_k^H \mathbf{F}_{\textrm{RF}} \mathbf{f}_{\textrm{BB},j}|^2 + \sigma_k^2, \forall k.
\end{equation}

\begin{IEEEproof}
See Appendix B.
\end{IEEEproof}

By Proposition 2, function (\ref{eq:f_r}) can be further reformulated  as
\begin{equation}
f_q(\mathbf{F}_{\textrm{RF}}, \mathbf{F}_{\textrm{BB}}, \mathbf{r}, \mathbf{t}) = \sum_{k = 1}^{K} \log_2(1 + r_k) - \sum_{k = 1}^K r_k
+ \sum_{k = 1}^K (2 \sqrt{1 + r_k} \Re\{t_k^{*}\mathbf{h}_k^H \mathbf{F}_{\textrm{RF}} \mathbf{f}_{\textrm{BB},k}\} - |t_k|^2 \mathbf{C}_k),
\label{eq:f_q1}
\end{equation}
where $\mathbf{t} \triangleq [t_1, \ldots, t_K]^T$.
Now, with auxiliary variable vectors $\mathbf{r}$ and $\mathbf{t}$, the optimization problem (\ref{eq:optimization problem}) can be transformed as
\begin{equation}
\begin{aligned}
\label{eq:optimization problem1}
\{\mathbf{F}_\textrm{RF}^{\star}, \mathbf{F}_\textrm{BB}^{\star}, \mathbf{r}^{\star}, \mathbf{t}^{\star}\} &~=~\arg \max ~~ f_q(\mathbf{F}_{\textrm{RF}}, \mathbf{F}_{\textrm{BB}}, \mathbf{r}, \mathbf{t})\\
{\rm s.t.} ~~ &\mathbf{F}_\textrm{RF}(i,j) \in \{0,\mathcal{F}\}, \forall i,j, \\
&\|\mathbf{F}_\textrm{RF}(i,:)\|_0 = 1, \forall i, \\
&\|\mathbf{F}_\textrm{RF}\mathbf{F}_\textrm{BB}\|_F^2 = P.
\end{aligned}
\end{equation}
To efficiently solve this problem, we propose to iteratively update the variables $\mathbf{F}_{\rm RF}, \mathbf{F}_{\rm BB}, \mathbf{r}$, and $\mathbf{t}$ to find at each iteration the conditionally optimal solution of one variable matrix/vector given others.
While the conditionally optimal $\mathbf{r}$ and $\mathbf{t}$ are already shown in (\ref{eq:r_k}) and (\ref{eq:y_k}), respectively, in the following, we turn to develop the iterative design of the digital beamformer and analog beamformer with given optimal $\mathbf{r}^\star$ and $\mathbf{t}^\star$.

To facilitate the analog and digital beamformer design, we rewrite the objective function in (\ref{eq:optimization problem1}) as the following form:
\begin{equation}
f_q(\mathbf{F}_{\textrm{RF}}, \mathbf{F}_{\textrm{BB}}, \mathbf{r}, \mathbf{t}) = \sum_{k = 1}^{K} \log_2(1 + r_k) - \sum_{k = 1}^K r_k - \sum_{k = 1}^K |t_k|^2 \sigma_k^2 + \delta,
\label{eq:f_q2}
\end{equation}
where
\begin{eqnarray}
\delta &\triangleq &\sum_{k = 1}^K 2 \sqrt{1 + r_k}\Re \{t_k^{*}\mathbf{h}_k^H \mathbf{F}_{\textrm{RF}} \mathbf{f}_{\textrm{BB},k}\} - \sum_{k = 1}^K \sum_{j = 1}^K |t_j|^2|\mathbf{h}_j^H \mathbf{F}_{\textrm{RF}}\mathbf{f}_{\textrm{BB},k}|^2,\label{eq:delta}\\
~ & = &\sum_{k = 1}^K (\sqrt{1 + r_k}(t_k^{*}\mathbf{h}_k^H \mathbf{F}_{\textrm{RF}}\mathbf{f}_{\textrm{BB},k}
+ \mathbf{f}_{\textrm{BB},k}^H \mathbf{F}_{\textrm{RF}}^H \mathbf{h}_k t_k) -  \mathbf{f}_{\textrm{BB},k}^H \mathbf{A}\mathbf{f}_{\textrm{BB},k}),\label{eq:delta1}\\
\non
\end{eqnarray}
with
\begin{equation}
\mathbf{A} \triangleq \mathbf{F}_{\textrm{RF}}^H\sum_{j = 1}^K |t_j|^2 \mathbf{h}_j\mathbf{h}_j^H \mathbf{F}_{\textrm{RF}}.
\end{equation}
From the above equivalent transformation we can see, when $\mathbf{r}$ and $\mathbf{t}$ are all fixed, maximizing $f_q$ in (\ref{eq:optimization problem1}) is equivalent to maximizing the last term of $f_q$ in (\ref{eq:f_q2}), i.e. $\delta$ which is given by (\ref{eq:delta1}).
Roughly speaking, $\delta$  can be deemed as a summation of transformed SINRs of $K$ users, where the ratio form of each SINR is converted to the difference between useful power and multiuser interference.
Motivated by this finding, the conditionally optimal analog beamformer and digital beamformer can be determined by the following problem:
\begin{equation}
\begin{aligned}
\label{eq:sub_problem2}
\{\mathbf{F}_\textrm{RF}^{\star}, \mathbf{F}_\textrm{BB}^{\star}\} &~=~\arg \max ~~ \delta\\
{\rm s.t.} ~~ &\mathbf{F}_\textrm{RF}(i,j) \in \{0,\mathcal{F}\}, \forall i,j, \\
&\|\mathbf{F}_\textrm{RF}(i,:)\|_0 = 1, \forall i, \\
&\|\mathbf{F}_\textrm{RF}\mathbf{F}_\textrm{BB}\|_F^2 = P.
\end{aligned}
\end{equation}
In the following two subsections, we attempt to iteratively design the conditionally optimal analog beamformer and digital beamformer based on optimization problem (\ref{eq:sub_problem2}).

\subsection{FP-based Analog Beamformer Design}

When $\mathbf{r}$, $\mathbf{t}$, and the digital beamformer $\mathbf{F}_\textrm{BB}$ are fixed, the problem of conditionally optimal analog beamformer design can be presented as follows:
\begin{equation}
\begin{aligned}
\label{eq:sub_problem_frf}
\mathbf{F}_{\rm RF}^{\star} &~=~\arg \max ~~ \delta\\
{\rm s.t.} ~~ &\mathbf{F}_{\rm RF}(i,j) \in \{0,\mathcal{F}\}, \forall i,j, \\
&\|\mathbf{F}_{\rm RF}(i,:)\|_0 = 1, \forall i. \\
\end{aligned}
\end{equation}
In this analog beamformer design problem, we can ignore the power constraint since the digital beamformer can be adjusted to satisfy the power constraint.
However, this problem is still difficult to be solved due to the discrete phase values and the $l_0$ norm constraint on the analog beamformer. In an effort to seek an available solution, we first rewrite the analog beamformer in the following form
\begin{equation}
\mathbf{F}_{\textrm{RF}} \triangleq \mathbf{S} \mathbf{F}_{\textrm{set}},
\label{eq:f_rf}
\end{equation}
where $\mathbf{F}_{\textrm{set}}$ is defined as
\begin{equation}
 \mathbf{F}_{\textrm{set}} \triangleq \left[
 \begin{array}{cccc}
 \mathbf{f}_{\textrm{set}} & \mathbf{0} & \ldots & \mathbf{0} \\
 \mathbf{0} & \mathbf{f}_{\textrm{set}} &  & \mathbf{0} \\
 \vdots &   & \ddots & \vdots \\
 \mathbf{0} & \mathbf{0} & \ldots & \mathbf{f}_{\textrm{set}}
 \end{array}
 \right]_{N_\textrm{RF} 2^B \times N_\textrm{RF}}
 \end{equation}
with
\begin{equation}
\mathbf{f}_{\textrm{set}} \triangleq \frac{1}{\sqrt{N_\textrm{t}}}\left[1, e^{j\frac{2\pi}{2^B}}, \ldots, e^{j\frac{2\pi(2^B - 1)}{2^B}}\right]^T,
\end{equation}
which contains all possible phases of a PS.
$\mathbf{S} \in \{0,1\}^{N_\textrm{t} \times N_\textrm{RF} 2^B}$ is a binary matrix and $\mathbf{S}(i,j) = 1$ indicates that the $i$-th antenna is mapped to the $\lceil j/2^B \rceil$-th RF chain with corresponding value $\mathbf{F}_{\textrm{set}} (j,\lceil j/2^B \rceil)$, where $\lceil \cdot\rceil$ denotes the ceiling operation.
Therefore, with known $\mathbf{F}_\textrm{set}$, the analog beamformer design problem can be reformulated as
\begin{equation}
\begin{aligned}
\label{eq:sub_problem3}
\mathbf{S}^{\star}=& \arg ~\max ~~ \delta\\
\textrm{s.t.} ~~&\|\mathbf{S}(i,:)\|_0 = 1, \forall i,\\
&~\mathbf{S}(i,j) \in \{0,1\}, \forall i,j.
\end{aligned}
\end{equation}

The optimization problem (\ref{eq:sub_problem3}) is a typical 0/1 integer programming problem.
Although it is still non-convex due to the discrete values of $\mathbf{S}$, the optimal
$\mathbf{S}^{\star}$ can be found by some off-the-shelf solvers (e.g. Mosek optimization tools) using Branch and Bound methods \cite{BB}, and then the corresponding optimal analog beamformer $\mathbf{F}_{\textrm{RF}}^{\star}$ can be obtained by
\begin{equation}
\mathbf{F}_{\textrm{RF}}^{\star} = \mathbf{S}^{\star} \mathbf{F}_{\textrm{set}}.
\label{eq:f_rf1}
\end{equation}

\subsection{FP-based Digital Beamformer Design}

With given $\mathbf{r}, \mathbf{t}$, and analog beamformer $\mathbf{F}_\textrm{RF}$, the problem of the conditionally optimal digital beamformer design can be expressed as:
\begin{equation}
\begin{aligned}
\label{eq:sub_problem_fbb}
\mathbf{F}_\textrm{BB}^{\star} ~=~&\arg \max ~~ \delta\\
{\rm s.t.} ~~ &\|\mathbf{F}_\textrm{RF}\mathbf{F}_\textrm{BB}\|_F^2 = P.
\end{aligned}
\end{equation}
The optimal solution of problem (\ref{eq:sub_problem_fbb}) can be determined by Lagrangian multiplier method. By introducing a multiplier $\mu$ for the power constraint in (\ref{eq:sub_problem_fbb}), we can form a Lagrangian function as
\begin{equation}
L_\delta = \delta + \mu (P - \|\mathbf{F}_\mathrm{RF}\mathbf{F}_\mathrm{BB}\|_F^2)
= \delta + \mu (P - \sum_{k = 1}^K \|\mathbf{F}_\mathrm{RF}\mathbf{f}_{\mathrm{BB},k}\|_2^2).
\end{equation}
Therefore, the problem (\ref{eq:sub_problem_fbb}) can be reformulated as
\begin{equation}
\{\mathbf{F}_\mathrm{BB}^\star, \mu^\star\} = \arg \max ~~L_\delta.
\end{equation}
Then the optimal solution of each row of $\mathbf{F}_\mathrm{BB}$ can be determined by setting the partial derivative of $L_\delta$ with respect to $\mathbf{f}_{\textrm{BB},k}$ and $\mu$ to zero, i.e.
\begin{equation}
\left\{
             \begin{array}{l}
             \frac{\partial L_\delta}{\partial \mathbf{f}_{\mathrm{BB},k}} = \mathbf{0}, \forall k,\\
             \frac{\partial L_\delta}{\partial \mu} = 0,
             \end{array}
\right.
\end{equation}
which yields the optimal digital beamformer as
\begin{equation}
\mathbf{f}_{\textrm{BB},k}^\star = (\mathbf{A} + \mu^\star \mathbf{F}_{\textrm{RF}}^H\mathbf{F}_{\textrm{RF}})^{-1} \sqrt{(1 + r_k)}\mathbf{F}_{\textrm{RF}}^H\mathbf{h}_k t_k, \forall k,\\
\label{eq:optimalfbb}
\end{equation}
where the optimal multiplier $\mu^\star$ is introduced for the power constraint
and can be easily obtained by bisection search.

After having the methods to find the conditionally optimal auxiliary vectors $\mathbf{r}^{\star}$ and $\mathbf{t}^{\star}$, digital beamformer $\mathbf{F}_\textrm{BB}^{\star}$ and analog beamformer $\mathbf{F}_\textrm{RF}^{\star}$, the overall procedure of the proposed hybrid beamformer design is straightforward. With appropriate initial $\mathbf{F}_\textrm{RF}$ and $\mathbf{F}_\textrm{BB}$, we iteratively update $\mathbf{r}$, $\mathbf{t}$, $\mathbf{F}_\textrm{RF}$, and $\mathbf{F}_\textrm{BB}$, until the convergence is observed.
For clarity, the proposed FP-based hybrid baemformer design algorithm is summarized in Algorithm 1.

\begin{algorithm}[htb]
\caption{FP-Based Hybrid Beamformer Design Algorithm}
\label{alg:AA}
\begin{small}
    \begin{algorithmic}[1]
    \REQUIRE $\mathbf{h}_1, \ldots, \mathbf{h}_k, B, N_\textrm{t}, N_\textrm{RF}, K$.
    \ENSURE $\mathbf{F}_{\textrm{RF}}^{\star}$, $\mathbf{F}_{\textrm{BB}}^{\star}$.
        \STATE {Initialize $\mathbf{F}_{\textrm{RF}}$ and $\mathbf{F}_{\textrm{BB}}$.}
        \WHILE {no convergence of $\mathbf{F}_{\textrm{RF}}$ and $\mathbf{F}_{\textrm{BB}}$ }
            \STATE {Update $\mathbf{r}^{\star}$ by (\ref{eq:r_k}).}
            \STATE {Update $\mathbf{t}^{\star}$ by (\ref{eq:y_k}).}
            \STATE {Update $\mathbf{F}_{\textrm{RF}}^\star$ by solving problem (\ref{eq:sub_problem3}) to obtain binary matrix $\mathbf{S}^{\star}$ and constructing as (\ref{eq:f_rf1}).}
            \STATE {Update $\mathbf{F}_{\textrm{BB}}^{\star}$ by (\ref{eq:optimalfbb}).}
        \ENDWHILE
        \STATE {Return $\mathbf{F}_{\textrm{RF}}^{\star}$ and $\mathbf{F}_{\textrm{BB}}^{\star}$.}
    \end{algorithmic}
    \end{small}
\end{algorithm}

\vspace{-1.0 cm}

\subsection{Convergence Analysis}
In this subsection, we will provide the proof of convergence of the proposed FP-based hybrid precoder design algorithm.
We start with introducing two inequations in the following proposition, which will be the foundation of the proof.

\vspace{0.3 cm}

\nid\textbf{Proposition 3.}
Let $R(\mathbf{F}_\mathrm{RF}, \mathbf{F}_\mathrm{BB})$  be the sum-rate (\ref{eq:sum-rate}) as a function of beamformers $\mathbf{F}_\mathrm{RF}$ and $\mathbf{F}_\mathrm{BB}$, $f_r(\mathbf{F}_\mathrm{RF}, \mathbf{F}_\mathrm{BB}, \mathbf{r})$ be the transformed objective function (\ref{eq:f_r}) by Proposition 1. Then, we will have
\begin{equation}
    R(\mathbf{F}_\mathrm{RF}, \mathbf{F}_\mathrm{BB}) \ge f_r(\mathbf{F}_\mathrm{RF}, \mathbf{F}_\mathrm{BB}, \mathbf{r}),
\end{equation}
where the equality holds if and only if each element of $\mathbf{r}$ satisfies (\ref{eq:r_k}).
Similarly, let $ f_q(\mathbf{F}_\mathrm{RF}, \mathbf{F}_\mathrm{BB}, \mathbf{r}, \mathbf{t})$ be the transformed objective function (\ref{eq:f_q1}) by Proposition 2. Then, we will have
 \begin{equation}
    f_r(\mathbf{F}_\mathrm{RF}, \mathbf{F}_\mathrm{BB}, \mathbf{r}) \ge f_q(\mathbf{F}_\mathrm{RF}, \mathbf{F}_\mathrm{BB}, \mathbf{r}, \mathbf{t}),
 \end{equation}
where the equality holds if and only if each element of $\mathbf{t}$ satisfies (\ref{eq:y_k}).

\vspace{0.3 cm}

\textit{Proof.}
These two inequations are straightforward since the values of $\mathbf{r}$ in (\ref{eq:r_k}) and $\mathbf{t}$ in (\ref{eq:y_k}) are all global optimal when the other variables are fixed.
\hfill$\blacksquare$

\vspace{0.3 cm}

Let $\mathbf{F}_\mathrm{RF}^{(n)}$, $\mathbf{F}_\mathrm{BB}^{(n)}$, $\mathbf{r}^{(n)}$, and $\mathbf{t}^{(n)}$ be the solutions of the $n$-th iteration obtained by Sec. III-A, (\ref{eq:optimalfbb}), (\ref{eq:r_k}), and (\ref{eq:y_k}), respectively. According to the iterative procedure of Algorithm 1, we emphasize that $\mathbf{r}^{(n)}$ is obtained by (\ref{eq:r_k}) using beamformers of the $n$-th iteration (i.e. $\mathbf{F}_\mathrm{RF}^{(n)}, \mathbf{F}_\mathrm{BB}^{(n)}$) and $\mathbf{t}^{(n)}$ is obtained by (\ref{eq:y_k}) with respect to beamformers and $\mathbf{r}$ of the $n$-th iteration (i.e. $\mathbf{F}_\mathrm{RF}^{(n)}, \mathbf{F}_\mathrm{BB}^{(n)}$, and $\mathbf{r}^{(n)}$).
Moreover, we define $\mathbf{t}^{(n+1,n)}$ as the result obtained by (14) based on the beamformers of the $(n+1)$-th iteration, but $\mathbf{r}$ of the $n$-th iteration (i.e. $\mathbf{F}_\mathrm{RF}^{(n+1)}, \mathbf{F}_\mathrm{BB}^{(n+1)}$, and $\mathbf{r}^{(n)}$).
Then by Proposition 3, we have
\begin{equation}
 \begin{aligned}
 R(\mathbf{F}_\mathrm{RF}^{(n+1)}, \mathbf{F}_\mathrm{BB}^{(n+1)}) &= f_r(\mathbf{F}_\mathrm{RF}^{(n+1)}, \mathbf{F}_\mathrm{BB}^{(n+1)}, \mathbf{r}^{(n+1)})
 \ge  f_r(\mathbf{F}_\mathrm{RF}^{(n+1)}, \mathbf{F}_\mathrm{BB}^{(n+1)}, \mathbf{r}^{(n)})\\
 &= f_q(\mathbf{F}_\mathrm{RF}^{(n+1)}, \mathbf{F}_\mathrm{BB}^{(n+1)}, \mathbf{r}^{(n)}, \mathbf{t}^{(n+1,n)})
 \ge f_q(\mathbf{F}_\mathrm{RF}^{(n+1)}, \mathbf{F}_\mathrm{BB}^{(n+1)}, \mathbf{r}^{(n)}, \mathbf{t}^{(n)}).
 \end{aligned}
\end{equation}
Assume $\mathbf{F}_\mathrm{RF}^{(n+1)}$ is the optimal solution using digital beamformer and two auxiliary vectors $\mathbf{r}, \mathbf{t}$ of the $n$-th iteration (i.e. $\mathbf{F}_\mathrm{BB}^{(n)}, \mathbf{r}^{(n)}$, and $\mathbf{t}^{(n)}$).
Similarly, set $\mathbf{F}_\mathrm{BB}^{(n+1)}$ as the optimal solution with
analog beamformer of the $(n+1)$-th iteration but $\mathbf{r}, \mathbf{t}$ of the $n$-th iteration (i.e. $\mathbf{F}_\mathrm{RF}^{(n+1)}$, and $\mathbf{r}^{(n)}, \mathbf{t}^{(n)}$).
We further have
\begin{equation}
 f_q(\mathbf{F}_\mathrm{RF}^{(n+1)}, \mathbf{F}_\mathrm{BB}^{(n+1)}, \mathbf{r}^{(n)}, \mathbf{t}^{(n)}) \ge f_q(\mathbf{F}_\mathrm{RF}^{(n+1)}, \mathbf{F}_\mathrm{BB}^{(n)}, \mathbf{r}^{(n)}, \mathbf{t}^{(n)})
  \ge f_q(\mathbf{F}_\mathrm{RF}^{(n)}, \mathbf{F}_\mathrm{BB}^{(n)}, \mathbf{r}^{(n)}, \mathbf{t}^{(n)}).
\end{equation}
Still by Proposition 3, we have
\begin{equation}
f_q(\mathbf{F}_\mathrm{RF}^{(n)}, \mathbf{F}_\mathrm{BB}^{(n)}, \mathbf{r}^{(n)}, \mathbf{t}^{(n)}) =  f_r(\mathbf{F}_\mathrm{RF}^{(n)}, \mathbf{F}_\mathrm{BB}^{(n)}, \mathbf{r}^{(n)})
=  R(\mathbf{F}_\mathrm{RF}^{(n)}, \mathbf{F}_\mathrm{BB}^{(n)}).
\end{equation}
Therefore, we can conclude that the sum-rate objective is monotonically non-decreasing after each iteration, i.e.
\begin{equation}
R(\mathbf{F}_\mathrm{RF}^{(n+1)}, \mathbf{F}_\mathrm{BB}^{(n+1)}) \ge R(\mathbf{F}_\mathrm{RF}^{(n)}, \mathbf{F}_\mathrm{BB}^{(n)}),
\end{equation}
which guarantees the convergence of Algorithm 1.
Moreover, simulation results in Section V also further verify the convergence.
It is worth noting that the algorithm may finally converge to a local optimum of optimization problem (\ref{eq:optimization problem1}).

\vspace{-0.3 cm}
\subsection{Complexity Analysis}

Finally, we provide a brief analysis of the time complexity of the proposed FP-based hybrid precoder design algorithm. As shown in Algorithm \ref{alg:AA}, steps 5 and 6 have dominant computational cost in the proposed approach. Particularly, in each iteration, finding the  digital beamformer in step 6 requires about $\mathcal{O}(K N_\textrm{RF}^3)$ complexity, which is mainly caused by the matrix inversion operation.
Computing the analog beamformer in step 5 needs to solve an $N_\textrm{RF} 2^B$-dimensional integer programming optimization problem with $N_\textrm{t}$ variables by CVX, which has $\mathcal{O}(\sqrt{N_\textrm{t} N_\textrm{RF} 2^B}[(N_\textrm{RF} 2^B)^3 N_\textrm{t}^2 + (N_\textrm{RF} 2^B)^2 N_\textrm{t}^3])$ complexity \cite{CVX}.
As a result, the overall time complexity of the proposed algorithm is about $\mathcal{O}(N_\textrm{iter} (K N_\textrm{RF}^3 + (N_\textrm{t} N_\textrm{RF} 2^B)^{\frac{5}{2}}(N_\textrm{RF} 2^B + N_\textrm{t})))$, where $N_\textrm{iter}$ is the number of iterations.
Obviously, with larger number of transmit antennas, there will be substantial growth of the complexity of this algorithm.
Therefore, in the next section, we propose another simple heuristic joint analog and digital beamformer design algorithm to dramatically reduce the time complexity but maintain acceptable performance.

\section{Heuristic Hybrid Beamformer Design}
\label{sc:Heuristic Design}

Similar with the proposed FP-based method described in the previous section, we propose an iterative scheme to obtain the analog beamformer and digital beamformer.
To reduce the complexity of solving the analog beamformer, we attempt to successively design each nonzero element of the analog beamformer when the digital beamformer $\mathbf{F}_\textrm{BB}$ is fixed. Then, with the effective baseband channel, we solve the digital beamformer based on uplink-downlink duality.

\subsection{Heuristic Analog Beamformer Design}
For the analog beamformer design, we propose to directly consider the original objective when the digital beamformer is fixed. Since the transmit power constraint can be satisfied by adjusting the digital beamformer, this analog beamformer design problem can be formulated as
\begin{equation}
\begin{aligned}
\label{eq:sub_problem_h2}
\mathbf{F}_\textrm{RF}^{\star} &~=~\arg \max ~~ R\\
{\rm s.t.} ~~ &\mathbf{F}_\textrm{RF}(i,j) \in \{0,\mathcal{F}\}, \forall i,j, \\
&\|\mathbf{F}_\textrm{RF}(i,:)\|_0 = 1, \forall i. \\
\end{aligned}
\end{equation}

This problem is difficult to solve due to the complicated form of the objective function and non-convex constraints of the analog beamformer.
However, from the $l_0$ norm constraint we can learn there is only one nonzero element in each row of the analog beamformer, which means that the analog beamformer matrix is sparse.
This fact motivates us to sequentially consider the design for each nonzero element (i.e. each row) of the analog beamformer.

Given an inital value of analog beamformer $\mathbf{F}_\textrm{RF}$, we aim to successively update each row of the analog beamformer.
Particularly, for the design of $i$-th row of the analog beamformer, $i = 1, \ldots, N_{\rm t}$, we attempt to conditionally increase the sum-rate performance $R$ with fixed other rows of $\mathbf{F}_\textrm{RF}$, which can be expressed as
\begin{equation}
\mathbf{F}_\textrm{RF} = [\underbrace{\mathbf{F}_\textrm{RF}(1,:)^T, \ldots, \mathbf{F}_\textrm{RF}(i-1,:)^T}_{\textrm{fixed}}, \underbrace{\mathbf{F}_\textrm{RF}(i,:)^T}_{\textrm{to be updated}},
\underbrace{\mathbf{F}_\textrm{RF}(i+1,:)^T, \ldots, \mathbf{F}_\textrm{RF}(N_\textrm{t},:)^T}_{\textrm{fixed}}]^T.\label{eq:update operation}
\end{equation}
Therefore, the sub-problem to determine the $i$-th row of the analog beamformer can be expressed as
\begin{equation}
\begin{aligned}
\mathbf{F}_\textrm{RF}^\star(i,:) ~=~ &\arg \underset{\mathbf{F}_\textrm{RF}(i,:)}{\max} ~R\\
{\rm s.t.} ~~&\mathbf{F}_{\rm RF}(i,j) \in \{\mathcal{F},0\}, \forall j,\\
 &\|\mathbf{F}_\textrm{RF}(i,:)\|_0 = 1.\label{eq:subproblem}
\end{aligned}
\end{equation}
Benefiting from using low-resolution PSs, we can perform a low-complexity two-dimensional exhaustive search over $N_\textrm{RF}$ RF chains and $\mathcal{F}$ with $|\mathcal{F}| = 2^B$, which only requires complexity of $\mathcal{O}(N_\textrm{RF} 2^{B})$.
By successively solving the subproblem (\ref{eq:subproblem}) over all the transmit antennas (i.e. rows of $\mathbf{F}_\textrm{RF}$), the analog beamformer can be obtained.

\subsection{Digital Beamformer Design Based on Uplink-downlink Duality}
When the analog beamformer is determined, we can obtain the effective baseband channel $\mathbf{\widetilde{h}}_k, k = 1, \ldots, K,$ for each user as
\begin{equation}
\mathbf{\widetilde{h}}_k^{H} \triangleq \mathbf{h}_k^H \mathbf{F}_{\rm RF}^\star. \label{eq:effective_channel}
\end{equation}
Then, the digital beamformer should be designed based on the effective baseband channel to further enhance the sum-rate performance.
Similar problems have been intensively investigated for conventional fully-digital beamforming schemes.
Therefore, in this paper, we adopted classic uplink-downlink duality theorem \cite{Codreanu 2007} and cyclic self-SINR-maximization (CSSM) algorithm \cite{Qian ICC 2012} to solve the digital beamformer design problem.
The procedure is briefly described below.

Based on the theory of uplink-downlink SINR duality, the original downlink beamformer design problem can be solved by solving its dual uplink optimization problem first, which is given by
\begin{equation}
\begin{aligned}
\{\mathbf{f}_1, \ldots, \mathbf{f}_K, q_1, \ldots, q_K\}& = \arg \max \sum_{k = 1}^K \log(1 + \textrm{SINR}_k^\textrm{ul})\\
\textrm{s.t.} ~& \mathbf{f}_k^H\mathbf{f}_k = 1, \forall k,\\
& \sum_{k = 1}^K q_k \le P,
\label{eq:dual uplink optimization problem}
\end{aligned}
\end{equation}
where
\begin{equation}
\textrm{SINR}_k^\textrm{ul} = \frac{q_k\mathbf{f}_k^H\mathbf{\widetilde{h}}_k\mathbf{\widetilde{h}}_k^H\mathbf{f}_k}{\mathbf{f}_k^H(\sum_{j\ne k}q_j\mathbf{\widetilde{h}}_j\mathbf{\widetilde{h}}_j^H)\mathbf{f}_k + \sigma_k^2}, k = 1, \ldots, K,
\end{equation}
is the dual uplink SINR for the $k$-th user, $\mathbf{f}_k, k = 1, \ldots, K,$ are the normalized digital beamformer vectors and $q_k, k = 1, \ldots, K,$ are uplink powers.
Applying the CSSM algorithm, the dual uplink optimization problem (\ref{eq:dual uplink optimization problem}) can be solved by the following iterative procedure:

\begin{enumerate}[\textbf{Step} 1:]
\item Obtain the max-SINR beamformer vectors $\mathbf{f}_k, k = 1, \ldots, K,$ as
\begin{equation}
\mathbf{f}_k = \left(\sum_{j = 1, j \neq k}^K q_j\mathbf{\widetilde h}_j\mathbf{\widetilde h}_j^H + \sigma_k^2\mathbf{I}\right)^{-1} \mathbf{\widetilde{h}}_k, \forall k,
\label{eq:f_k}
\end{equation}
each of which is further normalized by
\begin{equation}
\mathbf{f}_k = \frac{\mathbf{f}_k}{\|\mathbf{f}_k\|}, \forall k.
\label{eq:f_norm}
\end{equation}
\item Adjust powers $q_k$ with water-filling algorithm as
\begin{equation}
q_k = \left(\frac{1}{\mu} - \frac{1}{\varepsilon_k}\right)^{+}, \forall k,
\label{eq:wf}
\end{equation}
where $\varepsilon_k \triangleq \mathbf{\widetilde h}_k^H (\sum_{j \neq k} q_j\mathbf{\widetilde h}_j\mathbf{\widetilde h}_j^H + \sigma_k^2\mathbf{I})^{-1}\mathbf{\widetilde h}_k$, $\mu$ satisfies the power constraint, $(x)^+ = \max\{x,0\}$.

\item Alternatively update the beamformer vectors and powers until the stop criterion is satisfied (i.e. the difference of total power $\sum_{k = 1}^K q_k$ between two consecutive iterations is smaller than a given threshold).
\end{enumerate}

Given the optimal digital beamformer vectors and powers for the dual uplink problem, the original downlink problem can be easily solved. Based on the uplink-downlink duality, with the same beamformer vectors
and power constraint, the downlink system can achieve the same SINRs as the uplink case by using the downlink powers $p_k, k = 1, \ldots, K$, as
\begin{equation}
p_k = \sum_{i = 1}^{K} (\mathbf{B}^T)^{-1}(k,i), \forall k,
\label{eq:p_k}
\end{equation}
where $\mathbf{B}$ is a $K \times K$ matrix and
\begin{equation}
\mathbf{B}(i,j) = \left\{
    \begin{array}{ll}
        -|\mathbf{\widetilde{h}}_i^H\mathbf{f}_j|^2, &i \ne j ,\\
        |\mathbf{\widetilde{h}}_i^H\mathbf{f}_i|/\textrm{SINR}_i^\textrm{ul},  &i = j.
    \end{array}
    \right.
    \label{eq:B}
\end{equation}
Thus, the energy-included digital beamformer for original downlink system can be given by
\begin{equation}
\mathbf{\widehat{f}}_{\textrm{BB},k} = \sqrt{p_k} \mathbf{f}_k, \forall k.
\label{eq:f_bb_dl}
\end{equation}
To guarantee the power constraint, the final digital beamformer is normalized by
\begin{equation}
\mathbf{F}_\textrm{BB}^\star = \frac{\sqrt{P}\mathbf{\widehat{F}}_\textrm{BB}}{\|\mathbf{F}_\textrm{RF}\mathbf{\widehat{F}}_\textrm{BB}\|_F}.
\label{eq:normalization}
\end{equation}
where $\mathbf{\widehat{F}}_{\textrm{BB}} = [\mathbf{\widehat{f}}_{\textrm{BB},1}, \ldots, \mathbf{\widehat{f}}_{\textrm{BB},K}]$.

Based on the description in Sec. IV-A and Sec. IV-B, the iterative procedure of hybrid beamformer design is now straightforward. After selecting appropriate initial $\mathbf{F}_\textrm{RF}$ and $\mathbf{F}_\textrm{BB}$, we propose to iteratively update the analog beamformer and the digital beamformer until the convergence is satisfied.
The complete procedure for this heuristic hybrid beamformer design algorithm is summarized in Algorithm \ref{alg:HA}.

\begin{algorithm}[!t]
\caption{Heuristic Hybrid Beamformer Design}
\label{alg:HA}
\begin{small}
    \begin{algorithmic}[1]
    \REQUIRE $\mathbf{h}_1, \ldots, \mathbf{h}_K, B, N_\textrm{t}, N_\textrm{RF}, K$.
    \ENSURE $\mathbf{F}_{\textrm{RF}}^{\star}$, $\mathbf{F}_{\textrm{BB}}^{\star}$.
        \STATE {Initialize $\mathbf{F}_\textrm{RF}$, $\mathbf{F}_\textrm{BB}$.}
        \WHILE {no convergence of $\mathbf{F}_\textrm{RF}$ and $\mathbf{F}_\textrm{BB}$}
            \STATE {\textbf{Analog beamformer design:}}
            \FOR {$i=1$ : $N_\textrm{t}$}
                \STATE{Update $\mathbf{F}_\textrm{RF}(i,:)$ by solving problem (\ref{eq:subproblem}).}
            \ENDFOR
            \STATE {\textbf{Digital beamformer design:}}
            \WHILE {no convergence of $\mathbf{f}_k$ and $q_k, \forall k$}
                \FOR {$k = 1, \ldots, K$}
                    \STATE {Update uplink beamformer $\mathbf{f}_k$ as (\ref{eq:f_k}) and (\ref{eq:f_norm}).}
                \ENDFOR
                \FOR {$k = 1, \ldots, K$}
                    \STATE {Update uplink power $q_k$ as (\ref{eq:wf}).}
                \ENDFOR
            \ENDWHILE
            \STATE {Compute downlink beamformer $\mathbf{F}_\textrm{BB}$ by (\ref{eq:p_k})-(\ref{eq:f_bb_dl})}.
            \STATE {Normalize $\mathbf{F}_\textrm{BB}$ as (\ref{eq:normalization}).}
        \ENDWHILE
        \STATE {Return $\mathbf{F}_\textrm{RF}$ and $\mathbf{F}_\textrm{BB}$.}
    \end{algorithmic}
    \end{small}
\end{algorithm}

\subsection{Complexity Analysis}

We also give a brief complexity analysis of the proposed heuristic hybrid beamformer design algorithm. In each updating step, running through a full iteration over all $N_\textrm{t}$ antennas (i.e. $N_\textrm{t}$ rows of $\mathbf{F}_\textrm{RF}$) to obtain the analog beamformer requires $\mathcal{O}(N_\textrm{t}^2 N_\text{RF}^3 2^{B})$. Calculating the digital beamformer requires $\mathcal{O}(N_1 K N_\textrm{RF}^3)$, where $N_1$ denotes the number of iterations for CSSM algorithm.
As a result, the overall complexity of the proposed algorithm is $\mathcal{O}(N_\textrm{iter} (N_\textrm{t}^2 N_\text{RF}^3 2^{B} + N_1 K N_\textrm{RF}^3))$, where $N_\textrm{iter}$ is the number of iterations for the proposed algorithm.

Now, we turn to compare the complexities of two proposed algorithms.
When the number of transmit antennas $N_\textrm{t}$ is much larger than the number of users $K$ and RF chains $N_\textrm{RF}$, which is usually the case, the complexity of the FP-based hybrid beamformer design algorithm can be simplified as
\begin{equation}
C_\textrm{FP} = \mathcal{O}(N_\textrm{iter} N_\textrm{t}^{\frac{7}{2}}(N_\textrm{RF} 2^B)^{\frac{5}{2}}).
\end{equation}
Similarly, the complexity of the heuristic hybrid beamformer design algorithm can be given by
\begin{equation}
C_\textrm{Heuristic} = \mathcal{O}(N_\textrm{iter} N_\textrm{t}^2 N_\textrm{RF}^3 2^{B}).
\end{equation}
Therefore, we can draw a conclusion that the latter algorithm can significantly reduce the complexity especially when the large-scale antenna array is adopted. Moreover, simulation results in the next section indicate that the latter algorithm has faster convergence speed than the former scheme, which further strengthens our conclusion.

\section{Simulation Results}
In this section, we present simulation results to demonstrate the sum-rate and energy efficiency (EE) performance of two proposed hybrid beamformer designs with dynamic subarrays and low-resolution PSs. In the considered MU-MISO system, the BS is equipped with totally $N_\textrm{t} = N_\textrm{x} \times N_\textrm{y} = 6 \times 6 = 36$ antennas and $N_\textrm{RF} = 3$ RF chains, where antenna spacing $d$ is $\lambda/2$ and $\lambda$ denotes the wavelength. Without loss of generality, we assume the number of users $K$ is equal to the number of RF chains, i.e. $N_\textrm{RF} = K$. In the simulation, the channel parameter is set as $L = 5$ paths. Furthermore, horizontal and vertical AoDs are uniformly distributed over $\left[-\frac{\pi}{2}, \frac{\pi}{2}\right]$ and $\left[-\frac{\pi}{4}, \frac{\pi}{4}\right]$, respectively.
The signal-to-noise-ratio (SNR) is defined as $\frac{P}{\sigma_k^2}$ with $\sigma_k^2 = 1, k = 1, \ldots, K$. Finally, simulation results are averaged over $10^5$ channel realizations.

\subsection{Sum-rate Performance}

\begin{figure}[!t]
\centering
  \includegraphics[width=3.6 in]{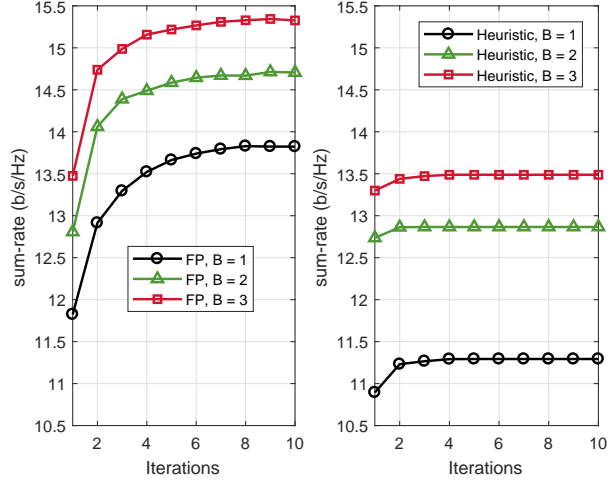}
  \vspace{-0.3 cm}
  \caption{The convergence of the proposed algorithms: Left: FP-based algorithm; Right: Heuristic algorithm (number of total antennas $N_\textrm{t} = 6 \times 6 = 36$, number of users $K = 3$, SNR = 10dB).}\label{fig:c_vs_iter}
  \vspace{-0.5 cm}
\end{figure}

In Fig. \ref{fig:c_vs_iter}, we first present the convergence of two proposed hybrid beamformer designs by plotting the sum-rate performance versus the number of iterations.
The SNR is fixed as 10dB.
Simulation results illustrate that the convergence of both two algorithms is really fast. Specifically, the FP-based hybrid beamformer design algorithm will converge within 8 iterations while the heuristic design algorithm has convergence within 4 iterations.
The convergence speed of the proposed two algorithms also indicates that the heuristic algorithm can effectively reduce the time complexity compared with the FP-based hybrid beamformer design algorithm even when a larger antenna array is used.
From Fig. \ref{fig:c_vs_iter}, we can also notice that the FP-based algorithm has better performance.
Moreover, with the growth of the resolution of PSs, the proposed two algorithms both tend to achieve better sum-rate performance.

\begin{figure}[!t]
\centering
  \includegraphics[width=3.6 in]{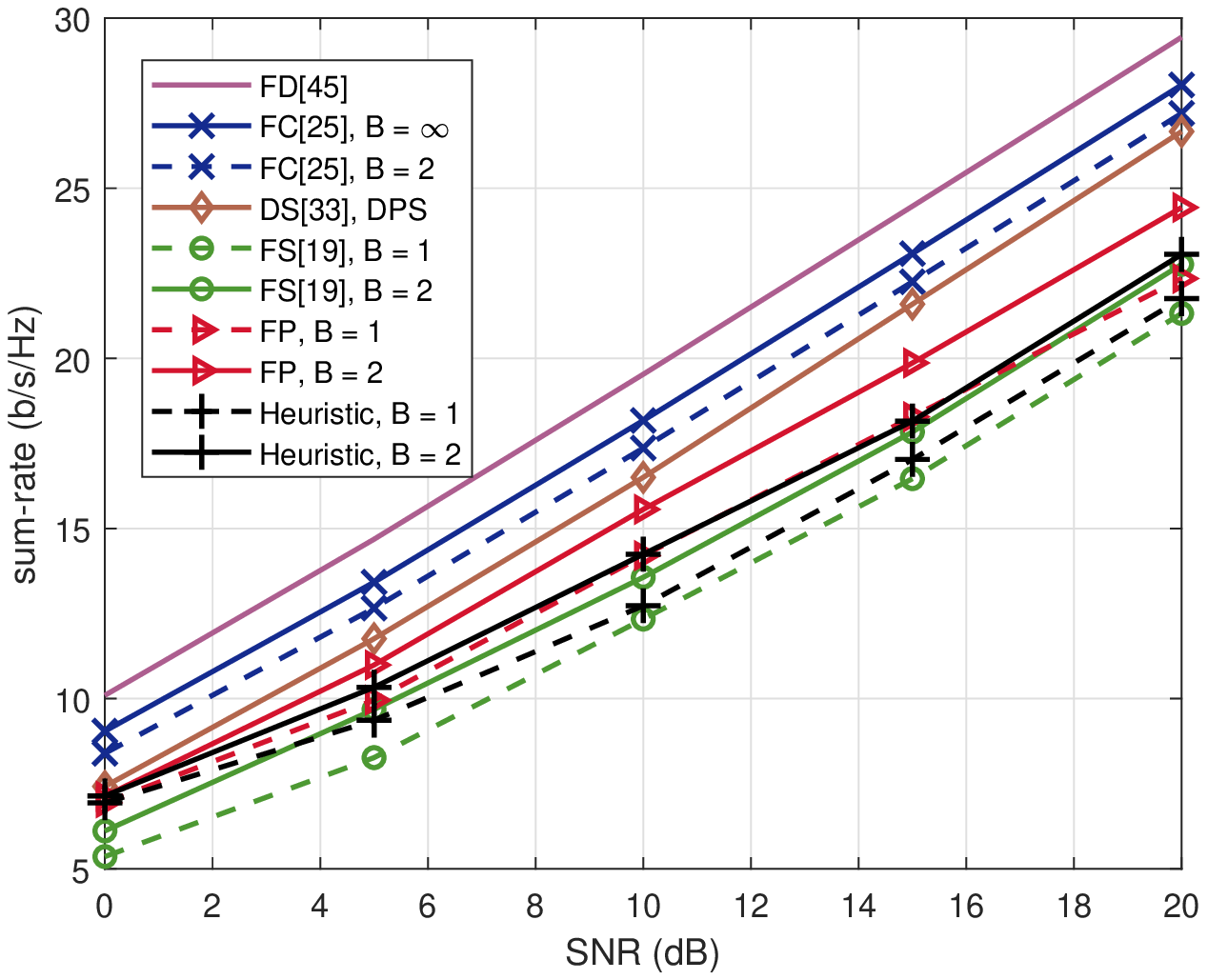}
  \vspace{-0.3 cm}
  \caption{Sum-rate versus SNR (number of total transmit antennas $N_\textrm{t} = 6 \times 6 = 36$, number of users $K = 3$).}
  \label{fig:c_vs_snr}
  \vspace{0.3 cm}
  \includegraphics[width=3.6 in]{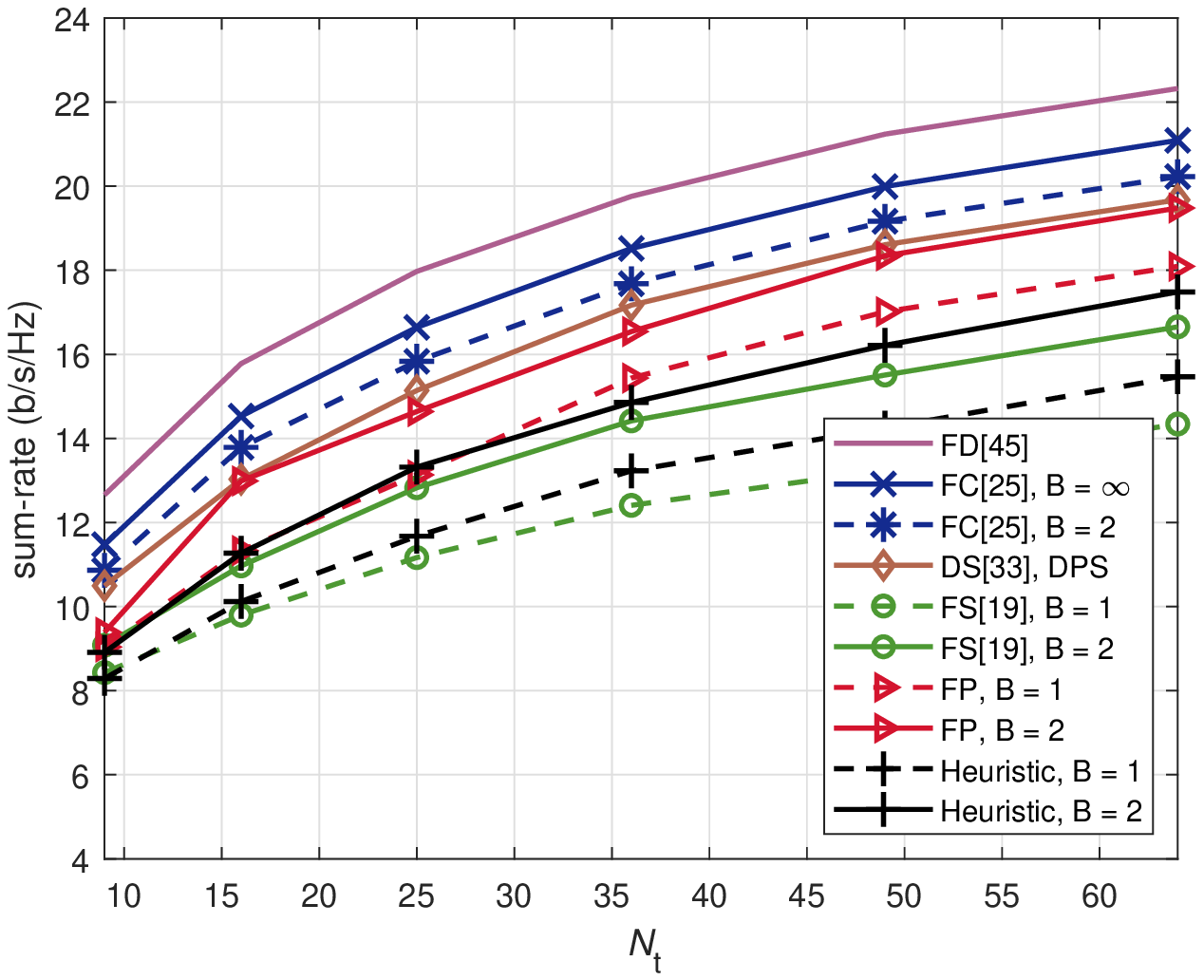}
  \vspace{-0.3 cm}
  \caption{Sum-rate versus number of total transmit antennas (number of users $K = 3$, SNR = 10dB).}
  \label{fig:c_vs_n}
   \vspace{-0.5 cm}
\end{figure}

Fig. \ref{fig:c_vs_snr} shows the achievable sum-rate versus SNR, in which we demonstrate the performance of two proposed hybrid beamformer design algorithms for the cases of using $B = 1, 2$-bit resolution PSs.
For the comparison purpose, we also include four algorithms: \textit{i}) Traditional fully-digital (\textbf{FD}) beamforming \cite{Qian ICC 2012}; \textit{ii}) fully-connected (\textbf{FC}) hybrid beamforming \cite{ZH J2018} for the case of using $B = 2$ bit and $B = \infty$ bit resolution PSs; \textit{iii)} dynamic subarray hybrid beamforming with double infinite-resolution PSs (\textbf{DS, DPS}) \cite{Yu SPAWC 2017}; \textit{iv}) fixed-subarray (\textbf{FS}) hybrid beamforming with successive interference cancelation (SIC) based beamformer design \cite{Gao JSAC 16}.
Since the original SIC-based beamformer design algorithm was developed for the infinite-resolution PS case, for the fair comparison purpose, we directly quantize the result of the analog beamformer to low-resolution values.
It can be observed from Fig. \ref{fig:c_vs_snr} that
the FP-based algorithm always achieves better performance than the heuristic scheme in different SNR ranges.
Specifically when $B = 2$, the FP-based algorithm can achieve satisfactory performance close to dynamic subarray scheme with double infinite-resolution PSs, which can serve as the performance upper bound of our algorithm because of the use of finer and more PSs.
More importantly, the proposed two hybrid beamforming solutions with dynamc subarrays and low-resolution PSs both can notably outperform the conventional fixed-subarray architecture.

In Fig. \ref{fig:c_vs_n}, we show the sum-rates of two proposed algorithms with respect to the number of antennas $N_\textrm{t}$.
In this case, we fix the number of users as $K = 3$ and SNR = 10dB.
A similar conclusion can be drawn from Fig. \ref{fig:c_vs_n} that the proposed algorithms both achieve superior performance compared with fixed-subarray schemes.
We can also expect that the sum-rate performance will become better as the increase of the number of antennas $N_\textrm{t}$, which can offer more antenna diversity and beamforming gain.

\begin{figure}[!t]
\centering
  \includegraphics[width=3.6 in]{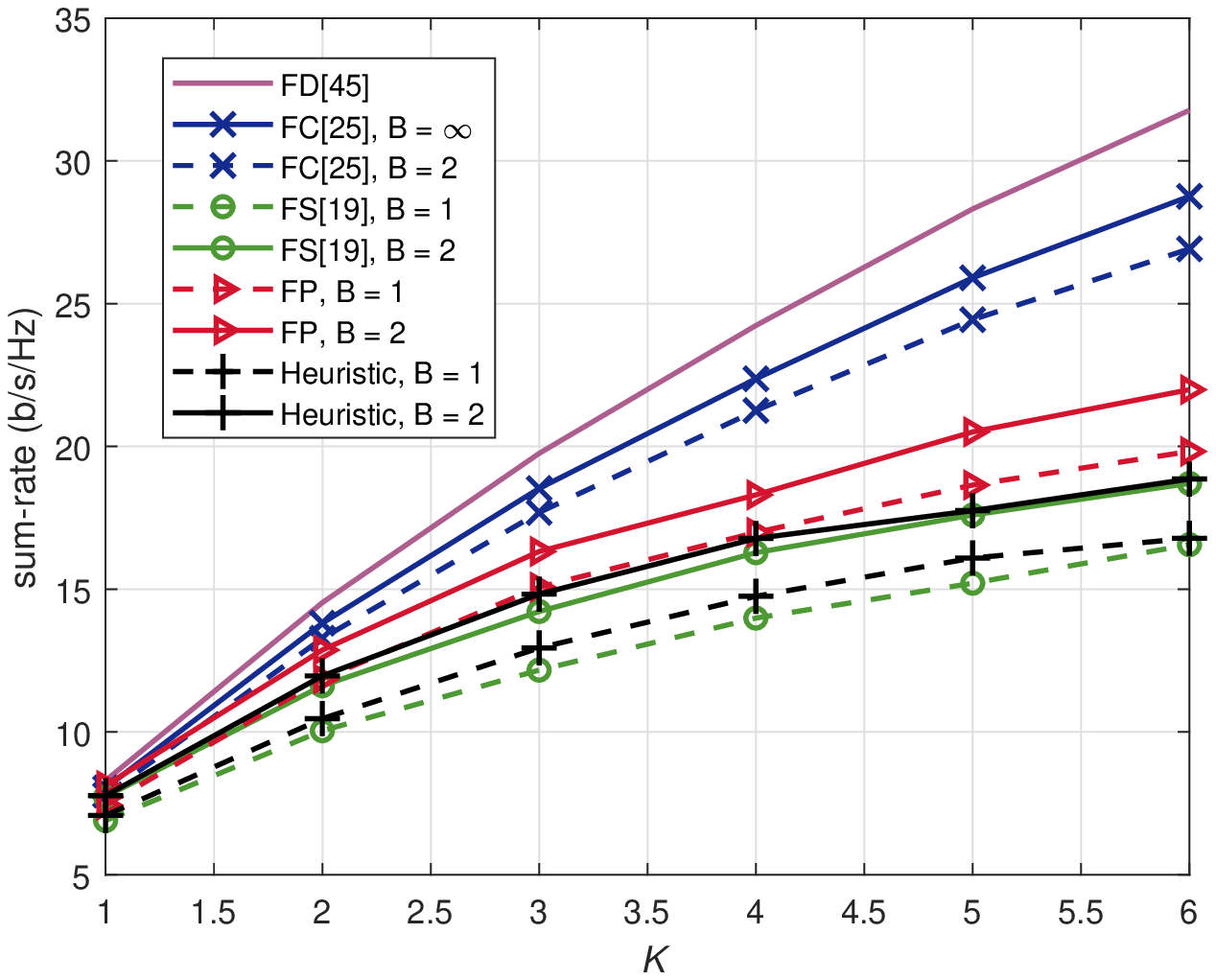}
  \vspace{-0.3 cm}
  \caption{Sum-rate versus number of users (number of total transmit antennas $N_\textrm{t} = 6 \times 6 = 36$, SNR = 10dB).}
  \label{fig:c_vs_k}
  \vspace{0.3 cm}
  \includegraphics[width=3.6 in]{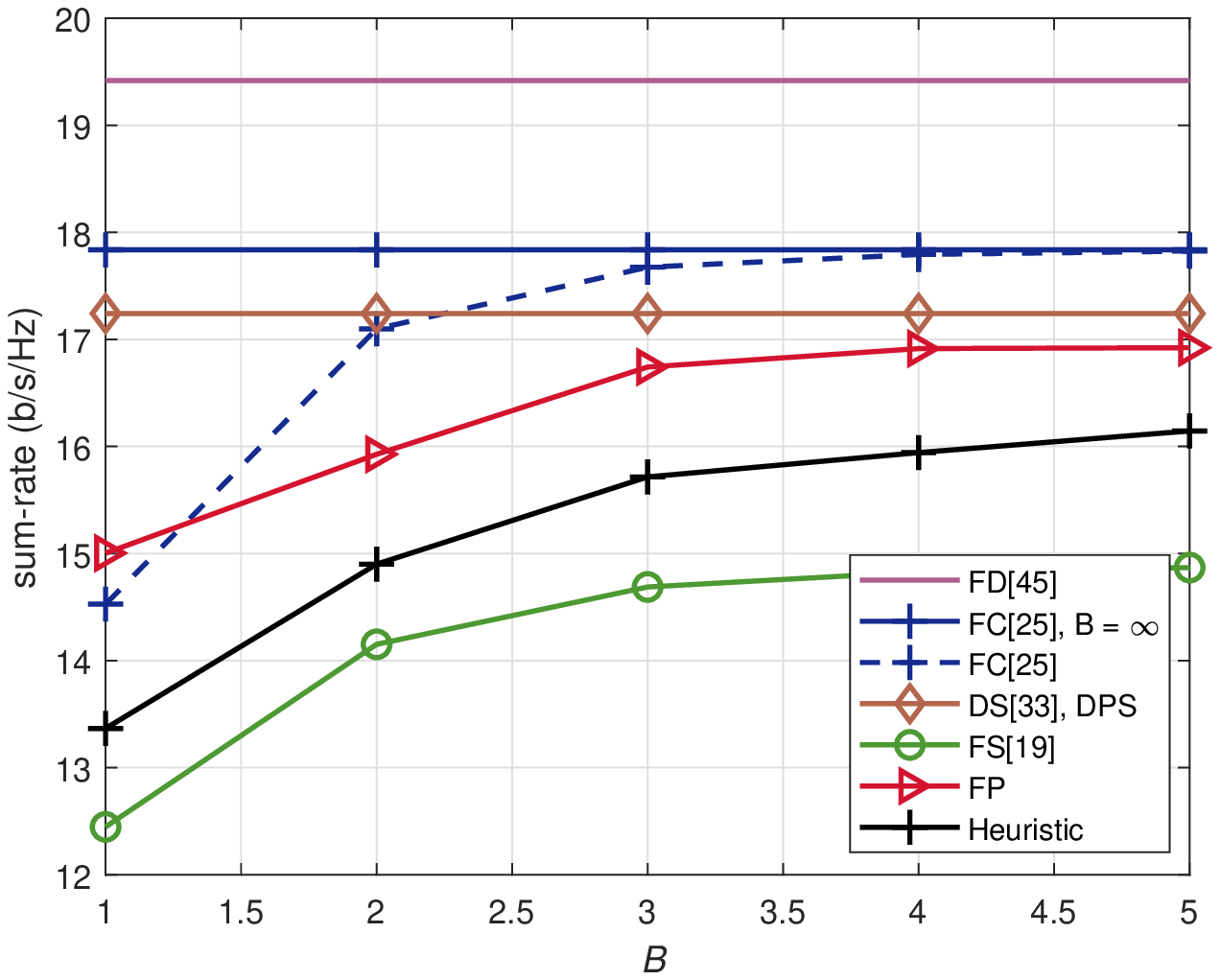}
  \vspace{-0.3 cm}
  \caption{Sum-rate versus the resolution of PSs (number of total antennas $N_\textrm{t} = 6 \times 6 = 36$, number of users $K = 3$, SNR = 10dB).}\label{fig:c_vs_b}\vspace{-0.7 cm}
\end{figure}

In Fig. \ref{fig:c_vs_k}, we illustrate the average sum-rate as a function of the number of users with the fixed number of antennas $N_\textrm{t} = 36$ and SNR = 10dB.
It further verifies that both two proposed algorithms always outperform fixed-subarray schemes with different user populations.
Furthermore, with the growth of the number of users, the performance gap between the fully-digital beamforming architecture and subarray schemes will become larger.
This phenomenon reveals an interesting fact: When the number of transmit antennas is fixed, serving more users will induce more multiuser diversity, but in the meanwhile, make it more difficult to eliminate the inter-user interference due to the less number of antennas in each subarray/user.
Therefore, it is important to properly set numbers of antennas and  users for subarray schemes.

Fig. \ref{fig:c_vs_b} shows the sum-rate performance as a function of $B$ to illustrate the influence of the resolution of PSs on the system performance. It can be clearly observed from Fig. \ref{fig:c_vs_b} that the sum-rate performance will first increase and then tend to saturate  with the growth of $B$.
When $B = 4$, the performance achieved by the FP-based algorithm can approach that of dynamic subarray scheme with double infinite-resolution PSs,
this result confirms that the low-resolution PSs are practical and sufficient for the real-world hybrid beamforming schemes.

\vspace{-0.8 cm}

\subsection{EE Performance}

In an effort to find the tradeoff between sum-rate performance and energy consumption, we illustrate the EE of the proposed designs, which is defined as

\begin{equation}
\eta \triangleq \frac{R}{P_{\textrm{tot}}}\label{eq:EE},
\end{equation}
where $P_\textrm{tot}$ is the total power consumption of the BS.
For the fully-digital beamforming architecture, the total power consumption is defined as
\begin{equation}
P_{\textrm{tot}}^\textrm{FD} \triangleq P + P_{\textrm{BB}} + N_\textrm{t}P_{\textrm{RF}},
\label{eq:P_fd}
\end{equation}
where $P$ is the transmit power, $P_{\textrm{BB}}$ and $P_{\textrm{RF}}$ are the powers consumed by the baseband processor and a RF chain, respectively.
For fully-connected hybrid beamforming architecture, the total power consumption can be introduced as
\begin{equation}
P_{\textrm{tot}}^\textrm{FC} \triangleq P + P_{\textrm{BB}} + N_\textrm{RF}P_{\textrm{RF}} + N_\textrm{t} N_\textrm{RF} P_{\textrm{PS}},
\label{eq:P_fc}
\end{equation}
where $P_{\textrm{PS}}$ is the energy consumed by a PS.
For traditional fixed-subarray schemes, the total power consumption is given by
\begin{equation}
P_{\textrm{tot}}^\textrm{FS} \triangleq P + P_{\textrm{BB}} + N_\textrm{RF}P_{\textrm{RF}} + N_\textrm{t} P_{\textrm{PS}}.
\label{eq:P_fs}
\end{equation}
Finally, for the proposed dynamic subarray scheme, the total power consumption can be written as
\begin{equation}
P_{\textrm{tot}}^\textrm{DS} \triangleq P + P_{\textrm{BB}} + N_\textrm{RF}P_{\textrm{RF}} + N_\textrm{t} P_{\textrm{PS}} + N_\textrm{t} P_{\textrm{SW}},
\label{eq:P}
\end{equation}
where $P_{\textrm{SW}}$ is the energy consumed by a SW.
In this simulation, the power consumptions of different devices in practical mmWave systems are listed in Table \ref{tab:p_c}.

\begin{table}[t]
\centering
\caption{Power Consumptions for Different Devices in mmWave Systems}
\vspace{-0.3 cm}
\label{tab:p_c}
\begin{tabular}{cc}
    \\[-2mm]
    \hline
    {  \small Hardware Device}&\qquad { \small Power Consumption (mW)}\\
    \hline
    \vspace{0mm}
    Baseband processor &\tabincell{l}{$P_{\textrm{BB}} = 200$} \\
    \vspace{0mm}
    RF chain &\tabincell{l}{$P_{\textrm{RF}} = 300$} \cite{Gao ICC 17} \\
    \vspace{0mm}
    PS &\tabincell{l}{$P_\textrm{PS} = 10 \, (B=1)$ \\$P_\textrm{PS} = 20 \, (B=2)$}\cite{Rial Access 16}\\
    \vspace{0mm}
    SW &\tabincell{l}{$P_{\textrm{SW}} = 5$} \cite{Gao ICC 17} \\
    \hline \vspace{-1.5 cm}
\end{tabular}
\end{table}

\begin{figure}[!t]
\centering
  \includegraphics[width=3.6 in]{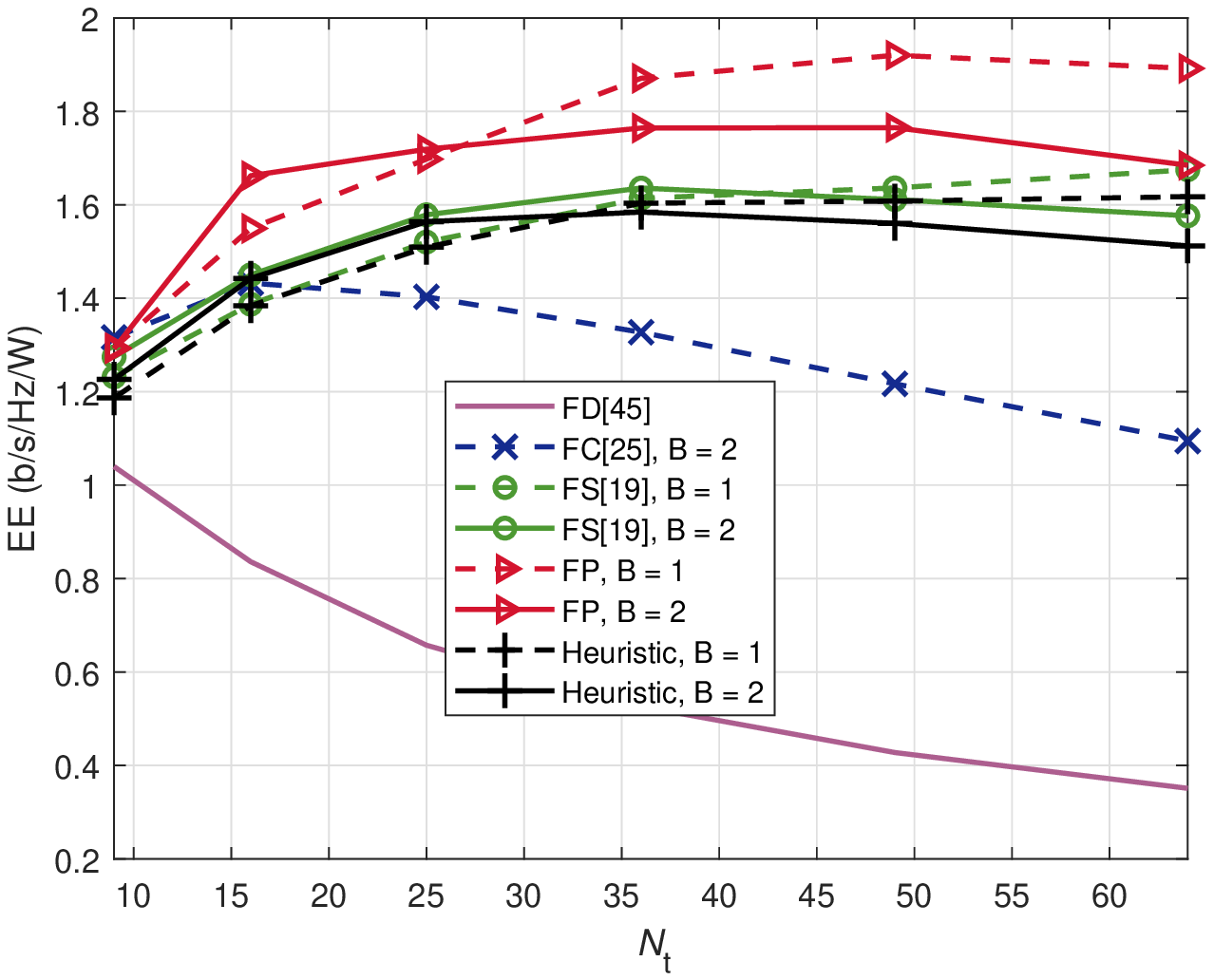}
  \vspace{-0.3 cm}
  \caption{EE versus number of total antennas $N_\textrm{t}$. (number of users $K = 3$, $P = 1W$).}\label{fig:E_vs_n}
  \vspace{0.5 cm}
  \includegraphics[width=3.6 in]{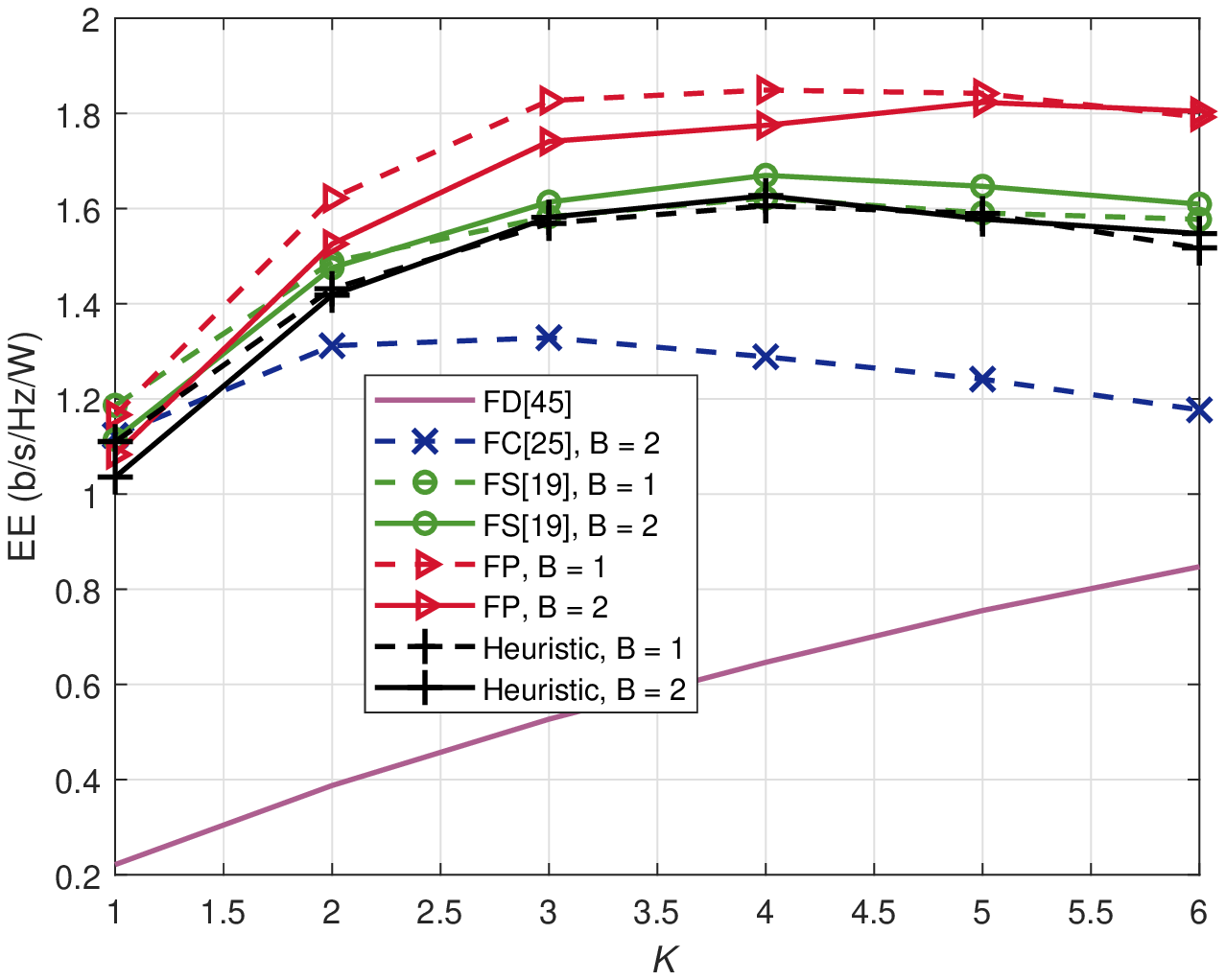}
  \vspace{-0.3 cm}
  \caption{EE versus K (number of total transmit antennas $N_\textrm{t} = 6 \times 6 = 36$, number of RF chains $N_\textrm{RF} = K$).}
  \label{fig:E_vs_k}\vspace{-0.0 cm}
\end{figure}

In Fig. \ref{fig:E_vs_n}, we present EE performance as a function of the number of transmit antennas $N_\textrm{t}$.
It can be observed in Fig. \ref{fig:E_vs_n} that the proposed FP-based hybrid beamformer design algorithm can maintain the best EE performance with the growing number of antennas.
Unfortunately, the EE achieved by the heuristic hybrid beamformer design algorithm is a little lower than the fixed-subarray scheme.
This is beacause that, even though the heuristic hybrid beamformer design has better sum-rate performance, it also requires an additional SW network to implement dynamic subarrays, which will consume more energy than the fixed-subarray scheme.
Then in Fig. \ref{fig:E_vs_k}, we further simulate the EE performance versus the number of users (i.e. number of RF chains). Similar conclusion as Fig. \ref{fig:E_vs_n} can be drawn. Moreover, the EE performance does not always grow with the increasing of number of RF chains and four or five RF chains can provide the best EE performance.

\begin{figure}
\centering
  \includegraphics[width=3.6 in]{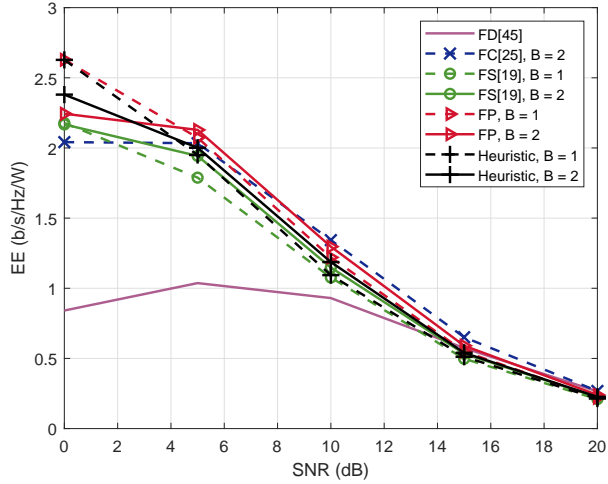}
  \vspace{-0.3 cm}
  \caption{EE versus SNR (number of total transmit antennas $N_\textrm{t} = 6 \times 6 = 36$, number of users $K = 3$).}
  \label{fig:E_vs_snr}\vspace{-0.5 cm}
\end{figure}

Finally, Fig. \ref{fig:E_vs_snr} illustrates the EE performance versus SNR when the numbers of transmit antennas and users are fixed as $N_\textrm{t} = 36$ and $K = 3$, respectively.
In this figure, the variance of the Gaussion noise is fixed as $\sigma_k^2 = 1, k = 1, \ldots, K$, and the transmit power $P$ is varying to simulate different SNR situations.
The proposed two algorithms can always achieve satisfactory performance than their competitors.
Moreover, with the growth of SNR, the EE gap among different schemes becomes smaller. The reason for this phenomenon is that the transmit power tends to dominate the total power consumption when SNR increases.

\vspace{-0.8 cm}

\section{Conclusions}
\label{sc:Conclusions}

In this paper, we introduced a novel efficient hybrid beamformer architecture with dynamic subarrays and low-resolution phase shifters (PSs) for a mmWave downlink multiuser multiple-input single-output (MU-MISO) system.
Aiming at optimizing the system sum-rate performance, we first proposed an iterative hybrid beamformer design algorithm based on the theory of fractional programming (FP). In order to reduce the time complexity, we also proposed another heuristic hybrid beamformer design algorithm which has low complexity even when the number of antennas got large.
The effectiveness of two proposed hybrid beamformer designs was validated by extensive simulation results, which illustrated that both two algorithms can remarkably outperform traditional fixed-subarray schemes.
For the future studies, it would be
interesting to extend the proposed hybrid beamformer architecture with
dynamic subarrays and low-resolution PSs to multi-user/multi-stream MIMO systems and wideband systems.

\begin{appendices}
\section{}
\begin{IEEEproof}[Proof of Proposition 1]
By introducing $K$ auxiliary variables $r_k, k = 1, \ldots, K$, to replace each ratio part (i.e. $\textrm{SINR}_k, k = 1, \ldots, K$) of objective in (\ref{eq:optimization problem}), the unconstrained sum-rate maximization problem can be rewritten as
\begin{equation}
\begin{aligned}
\max ~& \sum_{k = 1}^{K} \log(1 + r_k)\\
&\textrm{s.t.}~ r_k \le \textrm{SINR}_k, \forall k.
\label{eq:unconstrained problem}
\end{aligned}
\end{equation}
Obviously, (\ref{eq:unconstrained problem}) is a convex problem which can be easily solved by many optimization methods, e.g. Lagrangian multiplier method. By introducing $K$ multipliers $\lambda_k, k = 1, \ldots, K$, for each inequality constraint in (\ref{eq:unconstrained problem}), we can form a Lagrangian function as
\begin{equation}
L = \sum_{k = 1}^K \log(1 + r_k) - \sum_{k = 1}^K \lambda_k (r_k - \textrm{SINR}_k).\label{eq:L}
\end{equation}
The optimal $r_k^\star$ and $\lambda_k^\star, k = 1, \ldots, K,$ can be obtained by setting $\frac{\partial L}{\partial r_k} = 0$ and $\frac{\partial L}{\partial \lambda_k} = 0$:
\begin{eqnarray}
r_k^\star & = &\textrm{SINR}_k,  \forall k,\label{eq:r_k1}\\
\lambda_k^\star & = &\frac{1}{1 + r_k^\star},  \forall k. \label{eq:lambda_k}
\end{eqnarray}
With given $r_k^\star$ and $\lambda_k^\star$ in (\ref{eq:r_k1}) and (\ref{eq:lambda_k}), the maximal value of (\ref{eq:L}) can be written as
\begin{eqnarray}
L^\star &=&\sum_{k = 1}^K \log(1 + r_k^\star) - \sum_{k = 1}^K \frac{1}{1 + r_k^\star}(r_k^\star - \textrm{SINR}_k)\\
~ &=&\sum_{k = 1}^K \log(1 + r_k^\star) - \sum_{k = 1}^K \frac{r_k^\star}{1 + r_k^\star} + \sum_{k = 1}^K\frac{\textrm{SINR}_k}{1 + r_k^\star}\label{eq:opt_L}\\
~ &=& \sum_{k = 1}^K \log(1 + r_k^\star)
- \sum_{k = 1}^K \frac{r_k^\star(\sum_{j \ne k} |\mathbf{h}_k^H \mathbf{F}_{\textrm{RF}}\mathbf{f}_{{\rm BB},j}|^2 + \sigma_k^2)}{\sum_{j = 1}^K |\mathbf{h}_k^H \mathbf{F}_{\textrm{RF}}\mathbf{f}_{{\rm BB},j}|^2 + \sigma_k^2}+ \sum_{k = 1}^K\frac{|\mathbf{h}_k^H \mathbf{F}_{\textrm{RF}}\mathbf{f}_{{\rm BB},k}|^2}{\sum_{j = 1}^K |\mathbf{h}_k^H \mathbf{F}_{\textrm{RF}}\mathbf{f}_{{\rm BB},j}|^2 + \sigma_k^2} \\
~ &=& \sum_{k = 1}^K \log(1 + r_k^\star) - \sum_{k = 1}^K r_k^\star
+ \sum_{k = 1}^K \frac{|\mathbf{h}_k^H \mathbf{F}_{\textrm{RF}}\mathbf{f}_{{\rm BB},k}|^2(1 + r_k^\star)}{\sum_{j = 1}^K |\mathbf{h}_k^H \mathbf{F}_{\textrm{RF}}\mathbf{f}_{{\rm BB},j}|^2 + \sigma_k^2},\label{eq:opt_L1}\\
\non
\end{eqnarray}
which has the same form as $f_r$ in (10) in Proposition 1.

Conversely, when $\mathbf{F}_\textrm{RF}$ and $\mathbf{F}_\textrm{BB}$ are all fixed, it can be observed that (10) is a concave function of $\mathbf{r}$. Based on it, $\mathbf{r}$ can be directly determined by setting $\frac{\partial f_r}{\partial r_k} = 0, k = 1, \ldots, K$, i.e.
\begin{equation}
r_k^\star = \textrm{SINR}_k = \frac{|\mathbf{h}_k^H \mathbf{F}_{\textrm{RF}} \mathbf{f}_{\textrm{BB},k}|^2}{\sum_{j \ne k} |\mathbf{h}_k^H \mathbf{F}_{\textrm{RF}}\mathbf{f}_{\textrm{BB},j}|^2 + \sigma_k^2}, \forall k.
\end{equation}
Substituting the $\mathbf{r}^\star$ back into (\ref{eq:f_r}), the objective function (\ref{eq:f_r}) becomes that in (\ref{eq:optimization problem}) exactly, which completes the proof.
\end{IEEEproof}

\vspace{-1.0 cm}

\section{}
\begin{IEEEproof}[Proof of Proposition 2]
The function (\ref{eq:SINR_q}) can be rewritten as:
\begin{small}
\begin{eqnarray}
& ~ &\sqrt{1 + r_k}(t_k^*\mathbf{h}_k^H \mathbf{F}_{\textrm{RF}} \mathbf{f}_{\textrm{BB},k} + \mathbf{f}_{\textrm{BB},k}^H \mathbf{F}_{\textrm{RF}}^H \mathbf{h}_k t_k) - |t_k|^2 \left(\sum_{j = 1}^K |\mathbf{h}_k^H \mathbf{F}_{\textrm{RF}} \mathbf{f}_{\textrm{BB},j}|^2 + \sigma_k^2\right)\label{eq:1}\\
& = & \left(\sum_{j = 1}^K |\mathbf{h}_k^H \mathbf{F}_{\textrm{RF}} \mathbf{f}_{\textrm{BB},j}|^2 + \sigma_k^2\right) \left( \frac{\sqrt{1 + r_k}(t_k^*\mathbf{h}_k^H \mathbf{F}_{\textrm{RF}} \mathbf{f}_{\textrm{BB},k} + \mathbf{f}_{\textrm{BB},k}^H \mathbf{F}_{\textrm{RF}}^H \mathbf{h}_k t_k)}{\sum_{j = 1}^K |\mathbf{h}_k^H \mathbf{F}_{\textrm{RF}} \mathbf{f}_{\textrm{BB},j}|^2 + \sigma_k^2} -  |t_k|^2 \right)\label{eq:2}\\
& = &\left(\sum_{j = 1}^K |\mathbf{h}_k^H \mathbf{F}_{\textrm{RF}} \mathbf{f}_{\textrm{BB},j}|^2 + \sigma_k^2\right) \left(\frac{(1 + r_k)|\mathbf{h}_k^H \mathbf{F}_{\textrm{RF}} \mathbf{f}_{\textrm{BB},k}|^2}{(\sum_{j = 1}^K |\mathbf{h}_k^H \mathbf{F}_{\textrm{RF}} \mathbf{f}_{\textrm{BB},j}|^2 + \sigma_k^2)^2}-\left|\frac{\sqrt{1 + r_k}\mathbf{h}_k^H \mathbf{F}_{\textrm{RF}} \mathbf{f}_{\textrm{BB},k}}{\sum_{j = 1}^K |\mathbf{h}_k^H \mathbf{F}_{\textrm{RF}} \mathbf{f}_{\textrm{BB},j}|^2 + \sigma_k^2} - t_k\right|^2\right).\label{eq:3}
\end{eqnarray}
\end{small}
It can be verified from (\ref{eq:3}) that when the last absolute-norm term of (\ref{eq:3}) equals to zero, i.e.
\begin{equation}
t_k^{\star} = \frac{\sqrt{1 + r_k}\mathbf{h}_k^H \mathbf{F}_{\textrm{RF}} \mathbf{f}_{\textrm{BB},k}}{\sum_{j = 1}^K |\mathbf{h}_k^H \mathbf{F}_{\textrm{RF}}\mathbf{f}_{{\rm BB},j}|^2 + \sigma_k^2},
\end{equation}
function (\ref{eq:3}) has the same form as (\ref{eq:SINR_r}) exactly. Proposition 2 is therefore proved.
\end{IEEEproof}
\end{appendices}

\end{document}